\renewcommand{\deg}{^\circ}
\newcommand{\Msun}{{\rm M}_\odot}

\documentstyle[11pt,aaspp4]{article}

\begin{document}
\title{The 35-Day Evolution of the Hercules X-1 Pulse Profile:\\
Evidence For A Resolved Inner Disk Occultation of the Neutron Star}

\author{D. Matthew Scott \altaffilmark{1,2}, Denis A. Leahy \altaffilmark{3}, 
Robert B. Wilson \altaffilmark{2}} 
                                                                
\altaffiltext{1}{Universities Space Research Association}
\altaffiltext{2}{Space Science Directorate, SD-50, NASA/Marshall Space Flight
Center, Huntsville, AL 35812}
\altaffiltext{3}{Dept. of Physics, University of Calgary,
Calgary, Alberta, Canada T2N 1N4}                        
\authoremail{scott@gibson.msfc.nasa.gov,wilson@gibson.msfc.nasa.gov,
leahy@iras.ucalgary.ca}

\begin{abstract}
{\it Ginga} and Rossi X-ray Timing Explorer (RXTE) observations have allowed 
an unprecedented view of the recurrent systematic pulse shape changes 
associated with the 35-day cycle of Hercules X-1, a phenomenon currently unique 
among the known accretion-powered pulsars. We present observations of the 
pulse shape evolution. An explanation for the pulse evolution 
in terms of a freely precessing neutron star is reviewed and shown to have 
several major difficulties in explaining the observed pulse evolution pattern. 
Instead, we propose a phenomenological model for the pulse evolution based upon 
an occultation of the pulse emitting region by the tilted, inner edge of a 
precessing accretion disk. The systematic and repeating pulse shape changes 
require a {\it resolved} occultation of the pulse emission region. 
The observed pulse profile motivates the need for a pulsar beam consisting of 
a composite coaxial pencil and fan beam but the observed evolution pattern 
requires the fan beam to be focused around the neutron star and beamed in the 
antipodal direction. The spectral hardness of the pencil beam component 
suggests an origin at the magnetic polar cap, with the relatively softer fan 
beam emission produced by backscattering from within the accretion column, 
qualitatively consistent with several theoretical models for X-ray emission 
from the accretion column of an accreting neutron star.  
\end{abstract}

\keywords{pulsars: individual(Her X-1) --- X-rays: stars}

\section{Introduction}
Her X-1 is a 1.24 second period accretion-powered X-ray pulsar in a 
1.7 day circular orbit with a normal stellar companion, HZ Her. In addition 
to these basic periodicities this binary system has long been known to 
display an unusual 35-day long cycle of High and Low X-ray flux states. 
Within a single 35-day cycle are found a Main High and Short High X-ray
flux state lasting roughly ten and five days each respectively and separated
by ten day long Low states (see e.g. Scott \& Leahy 1999). Pulsations are 
detected during the High states but cease during the intervening Low states. 
The High states are punctuated by deep X-ray eclipses every 1.7 days 
indicating a line-of-sight close to the binary plane. 

A phenomenon currently known to exist only in Her X-1 is a repeating,
systematic evolution of the pulse profile that occurs during the 35-day 
cycle. Observations with the Large Area Counters 
on {\it Ginga} and the Proportional Counter Array (PCA) on RXTE have allowed 
an unprecedented view of the evolution of the pulse profile in shape and 
energy spectrum across both the Main and the Short High states. The 
{\it Ginga} observations cover the energy 
range 1-37 keV and sample five Main and two Short High states. They
are described in detail in Deeter et al. (1991), Scott (1993) and 
Deeter et al. (1998). The bulk of the data consists of a Main-Short-Main High
state sequence in April-May-June of 1989. Lightcurves and softness ratios for 
the 1988 and 1989 observations as well as some pulse profiles can be found in 
Leahy (1995a). The RXTE observations cover a consecutive Main and Short High 
state in September-October of 1996 and a Short-Main-Short High state sequence 
in September-October 1997. The 1996 Observations with the PCA in the range 
2-60 keV are presented in Scott et al. (1997a). 

In this paper, we present a simple phenomenological model for the pulse 
evolution based upon the occultation of the central X-ray source by the inner 
edge of a tilted, precessing, accretion disk. This choice is motivated by the 
observed association of pulse shape changes with {\it decreases} in overall 
X-ray flux near the end of the Main High state (see Scott 1993; Scott et al. 
1997a; Deeter et al. 1998; Joss et al. 1978 and Soong et al. 1990a) and the 
well known ability of a tilted, twisted, counter-precessing accretion disk to 
phenomenologically account for much of the complex 35-day optical and X-ray 
behavior displayed by the Her X-1/HZ Her system (Petterson 1975; 
Petterson 1977; Gerend \& Boynton 1976; Crosa \& Boynton 1980; 
Boynton, Crosa \& Deeter 1980; Middleditch 1983). 
We stress that most of the observed 35-day behavior has been associated with 
the {\it outer} portion of the accretion disk whereas we will demonstrate that
the pulse evolution must be an {\it inner} disk phenomenon.

Previous attempts to model the pulse shape evolution have relied on a
combination of neutron star free precession and obscuration by the accretion 
disk (Tr\"umper et. al. 1986; Kahabka 1987, 1989), obscuration 
by ``flaps'' of matter at the juncture of the accretion disk and pulsar 
magnetosphere (Petterson et al. 1991) and obscuration of the pulsar by a 
tilted precessing disk (Bai 1981; Averitsev et al. 1992). None of these 
previous models attempted to explain more than a few aspects of the pulse 
evolution. The model presented here refines the disk and pulsar beam geometry 
to qualitatively account for the observed pulse shape and its evolution during 
both the Main and Short High states.

A discussion of relevant aspects of the tilted, twisted, 
counter-precessing disk is presented in section 2. A  
summary of the main features observed in the pulse evolution during the
35-day cycle is presented in section 3. Section 4 discusses the absolute pulse 
phase alignment of the Main and Short High state pulses. Section 5 discusses 
the pulse evolution as a consequence of 
neutron free precession. Section 6 briefly discusses pulse evolution as
a consequence of changing mass accretion patterns onto the neutron star.
Section 7 presents a pulse evolution model based 
on an inner disk occultation and a simple X-ray beam configuration. In 
section 8 we discuss some implications of this inner disk occultation 
interpretation.

\section{A tilted, twisted and counter-precessing disk in Her X-1}
 
Her X-1 exhibits a rich variety of phenomena that appear to be well
explained by an accretion disk that is tilted, counter-precessing and 
twisted. We review some of the observational arguments for such a disk 
(see Priedhorsky \& Holt 1987 for an earlier review) 
and include some relevant new observations and interpretation. The occurrence 
of two distinct X-ray High states within the 35-day cycle has 
long been known. The Main High state was found soon after the discovery of 
Her X-1 (\cite{gia73}). The dimmer Short High state was first recognized 
in Copernicus observations (Fabian et al. 1973) and later in {\it Ariel 5} 
and {\it Uhuru} observations (Cooke \& Page 1975; Jones \& Forman 1976). 
Few extensive observations of the Short High state were made until
1989 with {\it Ginga} (Deeter et al. 1998). In figure 1 we show the 1-37 keV 
lightcurve obtained with {\it Ginga} in 1989 and for the August 1991 Main High 
state. The August 1991 observation caught a turn-on to the Main High state 
which is confirmed by simultaneous monitoring at lower sensitivity with the 
Burst and Transient Source Experiment (BATSE) on board the Compton Gamma Ray 
Observatory (CGRO). The regularity of occurrence of the Short High state
ten days after the end of the Main High state is now clearly demonstrated by
the ongoing monitoring of Her X-1 with the RXTE All Sky Monitor (ASM) 
(see figure 1; Scott \& Leahy 1999; and Shakura et al. 1999).

The companion star HZ Her is strongly heated by X-rays emanating from the 
neutron star and shows little variation in total magnitude, averaged over an
orbital cycle, throughout the 35-day cycle (Gerend \& Boynton 1976, 
Deeter et al. 1976). Optical pulsations produced by reprocessing of the 
primary X-ray flux have been observed during both the High and Low states 
(Middleditch \& Nelson 1976, Middleditch 1983). These observations show that 
total X-ray production at the neutron star is relatively constant, eliminating
gross periodic changes in accretion rate as the cause of the High-Low cycle. 
The strongest evidence for a tilted, counter-precessing disk in
Her X-1 comes from the complex set of systematic changes in the optical
orbital photometric light curve over the 35-day cycle that can be explained
by a combination of disk emission and disk shadowing/occultation of the
heated face of HZ Her (Gerend \& Boynton 1976).

The occurrence of two distinct X-ray High states within the 35-day cycle has 
been attributed to obscuration of the central X-ray source by a 
tilted, twisted, counter-precessing accretion disk (Petterson 1975; 
Petterson 1977). 
Such a disk geometry can be idealized as a set of tilted, concentric rings
in which the azimuth of the line-of-nodes of each successively smaller
ring shifts smoothly as one moves radially inwards (Petterson 1977).
The presence of {\it two} High states per 35-day cycle is naturally explained 
when the observer's line-of-sight lies close to the binary plane.    
The appearance of pulsations with significant cold matter absorption at
the onset of either High state (hereafter turn-on), is interpreted as the 
emergence of the pulsar from behind the tilted, outer rim of the accretion 
disk. The flux decline observed at the end of each High state shows little to
no absorption effects and begins as the line-of-sight to the pulsar is 
approached by the hot, tilted {\it inner} edge of the precessing accretion 
disk. This behavior is illustrated in figure 2 and compared to a succession 
of High states observed with {\it Ginga} in 1989 and with an averaged 35-day
lightcurve observed with the ASM on RXTE.

The direction of disk precession is apparently retrograde or
``counter-precessing''. The evidence for this comes from X-ray and optical
observations and theoretical considerations. X-ray absorption dips are 
observed in the Main High state just before eclipse that march toward earlier 
orbital phase as the High state progresses and at a frequency near but slightly 
lower than the sum of the orbital and 35-day frequencies (Crosa \& Boynton 
1980, Scott \& Leahy 1999). The optical lightcurve of HZ Her exihibits a 
systematic pattern of changes and a harmonic decomposition revealed that 
nearly all power is confined to a discrete set of frequencies composed of sums 
of the orbital and 35-day frequencies (Deeter et al. 1976). A uniformly 
counter-precessing disk will repeat the same disk-star orientation at the sum 
of the orbital and 35-day frequencies as will any phenomena dependent on the 
orientation. Prograde precession should cause phenomena to repeat at the 
difference of the orbital and 35-day frequencies but no such phenomena are 
observed. Theoretically, a tilted disk should also precess in a retrograde 
fashion (Katz 1973).                                         

\subsection{The outer disk and early High state behavior}  

X-ray pulsations appear during the flux rise at the start of the High 
states (i.e. the turn-on) accompanied by cold matter absorption in the X-ray 
spectrum (e.g. Parmar, Sanford \& Fabian 1980). 
The only Short High state turn-on that has been well observed 
to date is the May 1989 Short High state which is compared in figure 3
to the August 1991 Main High state turn-on. 
The turn-on's are nearly identical in form and both 
show an increase in softness ratio characteristic of cold matter absorption. 
The turn-on's are modeled in figure 3 by an X-ray point source emerging 
through an atmosphere with a gaussian density profile lying above the plane 
of a tilted, precessing disk. The disk angular velocity is $2\pi/34.85$ 
$\rm day^{-1}$, the scale height is $1/24$ the disk radius and the disk is 
tilted at $20\deg$ with respect to the orbital plane of Her X-1. The optical 
depth at the base of the disk atmosphere is 30. 
The disk model is described in more detail in section 7. The two turn-on's 
last about 3 hours in contrast to the eclipse egress which lasts only a few 
minutes (e.g. Leahy 1995b). The pulse profile exhibits no significant changes 
during the turn-on but merely increases in flux in different energy bands
(Deeter et al. 1998). The X-ray observations coupled with the optical 
observations imply that the primary cause of the High-Low flux cycle is 
obscuration. 

The flux after the beginning of the Main High state often shows a gradual 
increase by 20-50\% over the next 1-4 days. 
The pulse profile is relatively constant in shape during this period. We
propose that the X-ray flux rise is the result of viewing the neutron star 
through a hot dense lower corona lying just above the outer accretion disk 
surface. As the elevation of the observer's line-of-sight increases 
with respect to the nominal outer 
disk plane the amount of obscuration will decrease. This effect should also be 
present in the Short High state and in the system LMC X-4 where a tilted 
precessing disk is also postulated to occur 
(Lang et al. 1981). The existence of another 
much larger and lower density scattering corona is indicated by the existence 
of Low state flux at 5\% of the peak Main High state flux (e.g. Choi et al.
1997; Mihara et al. 1991). A two layer disk corona has been discussed 
theoretically by Schandl \& Meyer (1994). The lower corona has a temperature 
of $10^6$ K while the upper corona has a temperature of $10^{7.5}$ K. The 
observed $\sim50\% $ flux increase implies a column density of 
$1 \times 10^{24}$ $\rm cm^{-2}$ at the base of the lower corona for pure 
Thompson scattering. From figure 2, the duration of the flux increase implies 
a vertical angular thickness to the lower corona of $\approx 4\deg - 14\deg$ 
with respect to the nominal outer disk plane.  

\subsection{The inner disk and late High state behavior}  

If pure obscuration by a precessing, tilted, thin, and {\it planar} disk 
were the cause of the High-Low flux cycle then one might expect to 
observe: 1) rapid flux cutoffs at the ends of the High states equivalent to
the High state turn-on's 2) identical pulse profiles during the Main and 
Short High states and 3) nearly identical fluxes during the Main and Short
High states except for variations caused by geometric differences in coronal 
obscuration.
In contrast, both types of High state show gradual flux declines that last 
several days and without significant absorption effects. A tilted and 
{\it twisted} disk can explain the gradual flux decline if the disk is twisted 
in the direction of precession such that the azimuthal angle between the outer 
and inner disk line of nodes is $>90\deg$. In a tilted and twisted disk the 
line-of-nodes and tilt of individual contiguous disk rings varies smoothly
as one moves from the outer to inner disk radii. The hot, inner region of
the disk gradually covers the X-ray emitting region during a transition
to a Low state. We illustrate this type of disk model in figure 2 and compare 
it to the lightcurve observed with {\it Ginga} and an average lightcurve from 
the RXTE/ASM. A tilt of $\theta_{tilt,outer}=20\deg$ for the outermost ring 
is determined from the assumption of an observer elevation of 
$\alpha_{obs} = 5\deg$ and the observed 35-day phase separation of 
$\Delta \psi \sim 0.58$ for the Main and Short High state turn-on's
\footnote{$\theta_{tilt,outer} = \frac{\alpha_{obs}}
{\sin (\pi (\Delta \psi - 0.5))}$}. The required 
outer disk tilt is independent of the outer disk thickness.
We note that the geometry of the disk model proposed in Schandl \& Meyer 
(1994) is inconsistent with the observations since it predicts
cold matter absorption at the start of the Main High state with a gradual flux
decline caused by a covering up of the neutron star by an inner disk corona 
and the same events, but in {\it opposite} sequence, during the Short High 
state (e.g. see their figure 12). However, the same sequence of events is 
observed in both the Main and Short High states.

The overall spectral changes during the Main High state were further 
explored by comparing the 20-50 keV pulsed flux average Main High state 
lightcurve observed with BATSE with a 2-12 keV flux average Main High
state observed with RXTE/ASM. Following the procedure described in Scott
\& Leahy (1999), the BATSE pulsed flux light curve over
the timespan MJD 49933 to MJD 50507, obtained from folded-on-board data 
(see Bildsten et al. 1997), was sorted into orbital phase ``0.2'' turn-on
Main High states or ``0.7'' orbital phase turn-on's and averaged. Seven Main 
High states were used to construct the average 0.2 turn-on Main High state 
lightcurve and eight for the average 0.7 turn-on Main High state. Likewise, 
for the 2-12 keV RXTE/ASM a similar sorting and folding was done to construct 
average light curves for the timespan MJD 50146 to MJD 50947 with 12 and 
10 Main High states averaged to form, respectively, the average 0.2 turn-on 
and 0.7 turn-on Main High state lightcurves. During the flux decline at the 
end of the Main High state, the 20-50 keV pulsed flux dropped to the 
level of the background flux more than one day preceding a similar drop in 
the 2-12 keV flux in both turn-on type Main High states. In figure 4 we 
compare the softness ratio formed by the two lightcurves. The turn-on at 
35-day phase 0.0 shows a rapid increase in softness ratio consistent with 
decreasing absorption. The pre-eclipse dips also show up as decreased softness 
consistent with absorption. However, the softness ratio shows a large 
{\it increase} during the flux decline near the end of the Main High state 
followed by a large decline. The softness ratio increase is incompatible with 
either absorption or an energy independent Thomson scattering of an unresolved 
point source by a corona or disk atmosphere (as in the Schandl \& Meyer 1994 
model) as the sole cause of the flux decline. Therefore the flux decline at 
the end of the Main High state cannot be the result of an occultation of a 
{\it point} source by either a cold or a hot disk edge. 

The peak Short High state X-ray flux is only 30\% that of the peak Main High 
state flux and exhibits a quite different pulse profile. Dramatic changes in 
the pulse profile are observed during the last few days of the 
Main High state and throughout the Short High state (see Deeter et al. 1998 
and next section). If an obscuring region causes the gradual flux decline 
and the density scale height was much larger than the linear size of the 
pulse  emitting region then minimal pulse shape changes would be observed
as well as little difference in pulse profile between the Main and Short High 
states. 

Two possible explanations for the High state flux declines, pulse shape 
evolution and the low Short High state flux are 1) Systematic changes in the 
X-ray beaming direction are occurring in addition to those caused by neutron 
star rotation and/or 2) The scale height of a precessing obscuring inner disk 
region is indeed comparable in size to the pulse emitting region. If 
explanation 1) is correct then we are observing a combination of an
obscuration of a point source causing the flux declines and beaming changes
that cause both pulse shape changes and the Main and Short High state flux
difference. This possibility has been advocated by 
Tr\"umper et al. (1986) and Kahabka (1989) using beam changes caused by free 
precession of the neutron star coupled with obscuration by a precessing disk. 
In 2) the flux declines, the Main and Short High state flux and pulse shape 
differences and the pulse shape changes are caused purely by progressive 
disk occultation of an extended source.
We will discuss both these explanations in more detail after reviewing the 
phenomenology of the pulse evolution and phase alignment of the Main and Short 
High state pulses.

\section{Phenomenology of the pulse evolution in Her X-1}

We now review the salient features of the pulse profiles and their evolution 
presented in figures 5 and 6 and documented in Scott (1993) and Deeter et al. 
(1998). The early 1-37 keV Main High state pulse profile consists of a large 
main pulse and a smaller interpulse superposed on an underlying weakly pulsed 
component (see figure 5 for profiles and nomenclature of specific pulse 
features). The main pulse consists of two unequal shoulders (or peaks) at 
energies below 5 keV, but at higher energies a third central peak appears that 
grows with energy and dominates the main pulse profile above $\sim 20$ keV. 
These energy dependent features of the pulse are well known (see e.g. 
Soong et al. 1990b).
We display a subset of the {\it Ginga} observations in figure 6 showing the
evolution of the pulse profile during the Main and Short High states displayed 
in figure 1. 
From {\it Ginga} and other observations, we propose that the Main 
High state pulse profile evolution consists of three basic events listed in 
order of decreasing duration: 
1) A deepening and widening of a ``gap'' in the underlying component near the 
pulse phase of the {\it preinterpulse minima} 
that begins early in the  Main High state and continues until the 
end (i.e. pulse phase $\sim0.3$ in figure 6, top panels). This phenomenon can 
also be observed in {\it Uhuru} and {\it HEAO 1} pulse profiles 
(Joss et al. 1978; Soong et al. 1990a) and in recent RXTE observations 
(Scott et al. 1997a). The well known quasi-sinusoidal pulse profile that 
``appears'' near the end of Main High state may simply be the uncovering of 
this already present gap due to the disappearance of the overlying main pulse. 
The quasi-sinusoidal profile is the last pulsed feature to disappear before the 
High states end.
2) The disappearance of the {\it leading} and then the {\it trailing shoulder}
of the main pulse over a roughly two day period that starts six to seven days 
after the Main High state turn-on. The main pulse shows relatively little
change before this point. The disappearance of the leading shoulder of the
main pulse near the end of the Main High state has been noted many
times previously (e.g. Soong et al. 1990a; Joss et al. 1978; Kahabka 1989;
Sheffer et al. 1992; Scott 1993; Scott et al. 1997a; Deeter et al. 1998).
3) A rapid decay and disappearance of the spectrally {\it hard central peak} 
of the main pulse was observed with {\it Ginga} over an $\sim12$ hour period 
that took place within the time span of the shoulder decay. A similar rapid
decay of the main pulse was observed by {\it HEAO 1} but at lower resolution 
(Soong et al. 1990a).
In the context of the occultation model described below, this 
pulse shape evolution pattern suggests the presence, respectively, 
of three pulse emitting regions of decreasing size each roughly centered on 
the pulsar.       

Two Short High states have been observed in enough detail to follow
the pulse evolution. These are the {\it Ginga} observations shown in figure 1 
and recent RXTE observations (see Scott et al. 1997). However with these two 
Short High state observations, combined with the fragmentary observations of 
other Short High states, we can describe the following evolution pattern.
As the Short High state commences, the pulse profile is quite different 
in shape and lower in flux by $\sim70\%$ relative to the Main High state
pulse. 
Both {\it Ginga} and {\it Exosat} observations of the Short High state profile 
reveal a {\it small hard peak} and a 
larger {\it soft peak} separated by $180\deg$ in pulse phase and superposed 
on a {\it quasi-sinusoid} (see figure 6). During {\it Exosat} observations, the
{\it small hard peak} was actually larger than the {\it soft peak} at energies
above $\sim13$ keV (Kahabka 1987, 1989). {\it Ginga} observations also 
showed the {\it small hard peak} increasing in amplitude relative to the 
{\it soft peak} with increasing energy, but not exceeding that of the 
{\it soft peak}. The RXTE observation of the November 1996 Short High state 
did not reveal the presence of the {\it small hard peak}.
The {\it Ginga} observations show the {\it small hard peak} disappearing 
within one day of the turn-on as did the earlier {\it Exosat} observation 
(Kahabka 1987). The {\it Ginga} and RXTE observations showed that the 
{\it soft peak} also declined in flux, but more slowly, and disappeared three 
to four days after turn-on. A narrowing in width occurred in both cases.   

The Short High state {\it soft peak} contains an even softer component on the 
trailing side of the peak indicated by a spectral 
softening at approximately pulse phase $0.6$ (see Fig. 5). The very soft 
component is apparent 
as long as the {\it soft peak} is present. 
The {\it Exosat} pulse profiles presented by Kahabka (1987, 1989) 
also show this very soft component on the trailing side of the {\it soft peak}.

The amplitude of the {\it quasi-sinusoid} first increases and 
then decreases as the May 1989 Short High state progresses, and is 
about $180\deg$ out of phase compared with the Main High state 
{\it quasi-sinusoid} (see section 4). A similar flux increase of the 
{\it quasi-sinusoid} can be seen in the Short High state profiles displayed 
in figure 4.3 of Kahabka (1987). The overall flux stays relatively constant 
for about four days following the Short High state turn-on, but this is 
due to a flux {\it increase\/} in the {\it quasi-sinusoid} just 
balancing a flux {\it decrease\/} in the {\it small hard peak} and the 
{\it soft peak}.  The overall flux of the {\it quasi-sinusoid} is roughly
half that of the Main High state {\it quasi-sinusoid}. As in the 
Main High state, the gap that defines the {\it quasi-sinusoid} shows a 
decrease in width and depth at increasing energies.    

In summary, the Main High state pulse evolution involves a decay preferentially
on the leading edge of the {\it main pulse} and {\it interpulse} that begins 
late in the High state. In addition, the formation and continuous slow 
evolution of an underlying {\it quasi-sinusoidal} profile may also be occurring.
The Short High state involves the narrowing, decay and disappearance at very
different rates of the two peaks in the profile superposed on an underlying 
{\it quasi-sinusoid}. Evolution of the {\it quasi-sinusoid} is much slower 
and involves only subtle changes in the profile. These changes are illustrated
in figure 6. Overall, comparison of the {\it Ginga} observations with other 
observations of the pulse evolution are consistent with a repeating, stable
and systematic pattern of change in the pulse profile. Future observations
are needed to explore the details and stability of the pulse
evolution during the Short High state and especially the Main High state flux 
decline.

\section{Pulse Phase Alignment of the Main and Short High state pulses}

The overall pulse evolution pattern can only be completely understood if the 
proper phase alignment of the Main and Short High state pulse profiles 
is known. To properly align the Main and Short High state profiles two 
methods might be tried: 1) extrapolating the pulse timing ephemeris between 
High states across the Low state where pulsations are unobservable or
2) matching pulse features that are common to each profile.                

Figure 5 displays a Main High state pulse profile, a closeup of the 
{\it interpulse} of the same Main High state profile, and a Short High state 
profile. Both profiles are taken from an early point in their respective 
High states when the effects of the pulse profile evolution are minimal. 
The two High state profiles were phase aligned using the pulse timing 
extrapolation given in Deeter et al. (1998), in which a pulse phase ephemeris
is extrapolated from the April and June 1989 Main High states into
the June 1989 Short High state. 

At first glance the Main and Short High state pulse profiles seem quite 
different but there are actually a number of features common to each profile. 
The bottom of each panel has a hardness ratio for the profile. Note that the 
hardness ratio of the Main High state profile shows a dip at pulse phase 0.6 
and a spectral hardening near pulse phase 1.0. The Short High state profile 
also possesses a similar soft dip and spectral hardening separated by 0.4 in 
pulse phase. 

The phase alignment suggested by the spectral features implies that the 
Main High state {\it interpulse} and the Short High state peak {\it soft peak} 
should be matched together.
Examination of figure 5 shows that these two features exhibit very similar 
shapes, fluxes and energy dependence. The primary differences between the
two features may simply be due to the different backgrounds upon which each
feature rests, since the underlying {\it quasi-sinusoid} is shifted by 0.5
in phase between the two High states. 

The spectral softening at pulse phase 0.6 in either High state profile is 
apparent in all the High states observed by {\it Ginga} with the exception of 
the anomalous June 1989 Main High state. The spectral softening is also readily 
apparent in both the Main and Short High state profiles of the {\it Exosat} 
observations (see Kahabka 1987, 1989). 
Energy dependent differencing of the {\it Ginga } pulse profiles was used to 
isolate this soft excess feature. A pulse profile in a high energy band 
was scaled until a fit to the surrounding pulse in a lower energy band was 
obtained and then the difference was taken. This process revealed that the 
soft excess feature is very similar in shape, flux and color in all the High 
states which suggests that it is the {\it same} feature in both the Main and 
Short High states (see Scott 1993). The presence of the same soft 
excess feature in both the Main High state {\it interpulse} and Short High 
state {\it soft peak} strongly supports the phase alignment of figure 5.        

The phase alignment also suggests that the {\it hard central peak} of the 
Main High state pulse profile and the {\it small hard peak} of the Short High 
state profile are corresponding features. There is a significant flux 
difference between these two features but both hard peaks increase in width 
and amplitude at higher energies relative to the surrounding pulse. The width 
increase of the {\it small hard peak} is only marginally observable in 
figure 5, but is readily apparent in Kahabka (1989) and in RXTE/PCA
observations. In the {\it Exosat} observation of the Short High state shown 
by Kahabka (1989), the {\it small hard peak} is actually larger than the 
{\it soft peak} at higher energies. 

We can now describe the following similarities and differences of
features in the Main and Short High state profiles based upon the phase
alignment of figure 5. The Main High state interpulse and the Short High
state {\it soft peak} are the {\it same} feature, as is the soft excess feature 
within each profile. The Short High state {\it small hard peak} is a greatly 
reduced version of the {\it hard central peak} of the Main High state. The 
large soft two peaked component underlying the {\it hard central peak} in the 
Main High state is absent in the Short High state profile. The 
{\it quasi-sinusoidal} profile of the Short High state is reduced in flux by 
about 50\% compared to the Main High state {\it quasi-sinusoid}
and shifted in phase by roughly 0.5. The Short High state profile
can be viewed as a modified version of the Main High state profile rather
than a completely distinct pulse.

The initial appearance and probable phase alignment of the Short High state 
pulse profile require an explanation of 1) The disappearance 
of the soft shoulders of the Main High state main pulse. 2) The large drop in 
flux of the Main High state hard central peak 3) The nearly equivalent 
amplitudes of the Main High state interpulse 
and the Short High state soft peak. 4) The $\sim50\%$ drop in flux 
and $\sim180\deg$ phase difference between the two High state 
quasi-sinusoidal profiles.                                                   
                                                                            
\section{Free precession of the neutron star}

Force-free precession of the neutron star with a fixed beam geometry
has been proposed as the cause for the pulse shape change
between the Main and Short High states by Tr\"umper et al. (1986) and was
later elaborated in Kahabka (1987), \"Ogelman \& Tr\"umper (1988) and Kahabka 
(1989). They noted that free precession alone cannot explain the overall 
High-Low cycle since the beams needed to model the observed Main High state 
pulse are too wide to disappear from view and cause a Low state.
For example, the unpulsed and quasi-sinusoidal components of the profile 
must be produced by unbeamed emission or very wide beams, so free precession 
cannot cause these pulse components to disappear at the end of a High state.
A model based solely on free precession also has great difficulty 
explaining the cold matter absorption seen during turn-on in both kinds of 
High states. Therefore, a periodic obscuration of the X-ray source by the 
accretion disk is also assumed to explain the disappearance of the pulses and 
the overall High-Low cycle. A composite model with a precessing disk acting 
as an occulting body and a freely precessing neutron star in which both cause
pulse shape changes has been proposed by Kahabka (1989). The period and phase 
of both the precessing disk and the freely precessing neutron star must be 
closely locked together to prevent longterm drift between the pulse evolution 
pattern and the High-Low intensity cycle (for example, the leading edge decay 
of the {\it main pulse} has only been observed at the end of the Main High 
state). The observed random walk in the phase of the Main High state poses 
some difficulties for phase locking since a freely precessing neutron star 
should be a relatively good clock (see Boynton, Crosa \& Deeter 1980; 
Tr\"umper et al. 1986 and Baykal et al. 1993). 

A spinning body can exhibit periodic, force-free precession if two of its three 
principal axes of inertia are unequal (i.e. a symmetric top 
with $I_1 = I_2 \not= I_3$). A review of the basic equations of free 
precession can be found in Bisnovatyi-Kogan, Mersov \& Sheffer (1990).
Free precession can cause major changes in the observed pulse profile 
that repeat over the course of successive precession cycles. 
The pulse profile at a given pulse phase changes due to the slow variation 
in the angle between the magnetic dipole axis and the observer's line-of-sight.
Beams emanating from the magnetic polar caps or from any fixed location
on the neutron star (other than the figure axis) will appear to a distant 
observer to ``wobble'' in rotational latitude over the precession cycle.
Either end of the dipole axis, when facing the observer, will oscillate
sinusoidally in latitude during one free precession cycle. 
The observed free precession light curve is 
highly dependent on the angle between the angular momentum axis and the
observer's line-of-sight and the emission beam geometry and will be symmetric 
about the precession phases of the latitude extrema in the dipole motion.
The sensitivity to the beam profile depends mostly on the beam width. 
A narrow beam produces pulse components that can appear
and disappear from view as the wobbling takes place while components 
from wider beams will merely show variations in amplitude.

Another potentially observable signature of free precession is a characteristic 
variation in the pulse period over the free precession cycle 
(Bisnovatyi-Kogan, Mersov \& Sheffer 1990; Bisnovatyi-Kogan \& Kahabka 
1993). Measurement of this effect was initially a major motivation for 
observing Her X-1 with {\it Ginga}. We can estimate the size of the period 
change expected between the Main and Short High state using equation 7 of 
Bisnovatyi-Kogan \& Kahabka (1993) to be $\sim 5.0 \times 10^{-7}$ s. 
However the known random walk in pulse frequency (see Bildsten et al. 1997;
Deeter 1981) 
also causes pulse period changes. The expected period change between the Main 
and Short High state is given by:
\begin{equation}
\left < \Delta P^2 \right >^{\frac{1}{2}} = P^2_0 (St)^{\frac{1}{2}}
\end{equation}    
and has a value of $\sim 8.0 \times 10^{-7}$ s for a noise strength
$S = 1.8 \pm 0.8 \times 10^{-19}$ $\rm Hz^2\ s^{-1}$ and $t = 17.5$ days. 
The expected period changes are thus comparable and would be difficult to 
separate unless many Main-Short-Main High state pulse period measurements 
could be made. Currently, the most observable potential manifestation of 
free precession is the pulse evolution observed between and within the High 
states.

The ability of free precession to account for the change in pulse profile 
between the Main and Short High states was tested by Kahabka (1987) using 
{\it Exosat} observations.
The Main High state pulse profile was modeled using a gaussian pencil beam 
for the {\it hard central peak} and a coaxial gaussian fan beam for the 
{\it soft leading} and {\it trailing peaks}. The {\it interpulse} is produced 
by an identical antipodal fan beam. The opening angle of the fan beam was 
determined to be about $45\deg$. We show a sequence of pulse profiles evolving 
over the 35-day cycle using the parameters determined by Kahabka (1987, 1989) 
in figure 7. The pulses occur one day apart over the 35-day cycle and do not 
include the unpulsed or quasi-sinusoidal component. The Main High state must 
be centered roughly about day zero with the Short High state occurring 18 days 
later. From the figure several features can be noted: (1) The {\it interpulse} 
and the {\it soft two peak component} are always visible. (2) The 
{\it interpulse} flux increases (or decreases) as the {\it main pulse} flux 
decreases (or increases). (3) No leading edge decay of either the {\it main 
pulse} or {\it interpulse} occurs. (4) The pulse profile evolves smoothly and 
slowly over the 35-day cycle, with no rapid profile changes. (5) The decrease 
in pulsed intensity is only about 10\%-20\% between the Main and Short High 
state. All of these features are inconsistent with the evolution actually 
observed as described in section 3. Related criticisms of free precession have 
been made by Bisnovatyi-Kogan, Mersov \& Sheffer (1990).
In Kahabka (1989), a precessing disk was assumed to cause the
leading edge decay of the {\it main pulse} during the termination of the Main 
High state with some contribution from free precession. However, this model 
has difficulty in explaining the actual Short High state profile and 
the subsequent disappearance of both peaks in the profile as noted in points
(1) and (2) above.

A freely precessing neutron star with approximately axisymmetric fan and 
pencil beams has several general problems in explaining the observed 
pulse evolution in Her X-1. (1) The pencil beam cannot disappear from view 
without the fan beam disappearing as well. The {\it small hard peak} of the 
Short High state appears to be the same pulse component as the
{\it hard central peak} of the Main High state (see section 4), but without 
the surrounding {\it soft two peaked component} that is presumably produced 
by a fan beam.
(2) The preferential decay of one side of a fan beam cannot occur due to
free precession. The two cuts across a fan beam, which result in a two-peaked
pulse component, should change in amplitude and width simultaneously as
precession occurs. (3) Rapid profile changes, such as those observed near the 
end of the Main High state or the beginning of the Short High state,
are difficult to explain unless the beam profile has sharp edges. For example,
the flux of the {\it hard central peak} dropped by more than 70\% in 6.7 
hours near the end of the April 1989 Main High state. If we assume that the 
{\it hard central peak} is produced by a pencil beam and that the precessional 
motion of the pulsar is responsible for its amplitude decrease, then we can 
estimate the {\it latitudinal} width of the pencil beam. 
We conservatively estimate that the {\it hard central peak} flux drops to zero 
in approximately 12 hours. In this period, the beam axis moves approximately 
through an angle $4\Phi \Delta t/P_{35}$ away from the observer's 
line-of-sight.  With $\Phi\approx 12.5\deg$, this produces a latitudinal 
width over this portion of the beam of about $0.7\deg$. We can also estimate 
the {\it longitudinal} width of this portion of the beam to be $\sim40\deg$ 
from the observed pulse shape as the beam sweeps {\it across} our
line-of-sight. The beam responsible for the {\it hard central peak} would 
have to be $\sim 60$ times broader in longitude than in latitude, a quite
improbable situation. 

To explain a rapid change in the pulse profile observed with {\it HEAO 1} 
near the end of a Main High state (Soong et al. 1987) with a free precession 
model, Shakura, Postnov \& Prokhorov (1998) have invoked a temporary transition
from a symmetric top to a triaxial neutron star caused by a ``quake'' 
during the Main High state flux decline. The quake induces a rapid shift in the
rotational latitude of the magnetic axis and thus induces rapid changes
in the observed pulse profile. This explanation may work for one case of
observed rapid pulse profile change but the rapid decline of the pulse
during the Main High state now appears to be a normal and repeating phenomenon
rather than an anomalous event (see Deeter et al. 1998). It seems rather 
improbable that neutron star quakes could be arranged to regularly occur 
during the 35-day phase of the Main High state flux decline but not to occur 
at other times. In addition, there appears to be no evidence for an expected 
phase shift caused by the quake based on {\it Ginga} and RXTE observations. In 
conclusion, we find little evidence supporting neutron star free precession as 
the cause of the pulse shape changes in Her X-1. 

\section{Other causes of systematically recurring pulse evolution}
If the neutron star is not undergoing free precession then the precessional
motion of the inner disk becomes the primary candidate causing pulse
evolution. Pulse shape changes might be caused by modulation of matter flow
onto the magnetic field lines by a changing aspect between the magnetosphere 
and a tilted, precessing accretion disk. The cycle of High and Low states
can be attributed to obscuration by the precessing accretion disk while the
pulse shape evolution results from variations in the accretion column
structure induced by the changing pattern of matter entry onto the magnetic
field lines. 
As a naive example of a possible variation, assume the accreting matter
attaches directly to a dipole magnetic field at a radius $R_m$ and then
travels along the field lines onto the magnetic poles. The angular
extent of the accretion cap will be given by $\sin^2(\theta_C) = R/R_m$ where
$R$ is the radius of the neutron star (Lamb, Pethick \& Pines 1973).
An increase in $R_m$ will cause a decrease in the accretion
column radius at the neutron star surface and concentrate the energy
released by the accretion into a smaller area. A decrease in $R_m$ will
have the opposite effect.
A full evaluation of the coupling between a tilted, precessing accretion disk,
the neutron star magnetosphere and the net effect on the pulsar beam involves
complex physics that has not been undertaken as yet and is beyond the scope of
this discussion.

It seems plausible that some degree of systematic pulse evolution should be
caused by a changing disk orientation, but we argue that this is unlikely 
to be the {\it primary} cause of the pulse shape changes observed in Her X-1.
Nonlinear changes in the flow rate and pattern could be responsible for
rapid changes observed in the pulse profile, but these rapid changes
must be scheduled to occur just as the overall flux declines at the end
of the Main High state and not at other times. 
This mechanism should also cause both {\it increases} and {\it decreases} in 
the pulse profile. In the visible High states only decreases in pulse features 
are ever observed.\footnote{During the
Short High state the {\it small hard peak} and
the {\it soft peak} disappear, but this is accompanied by a compensating
small flux increase in the quasi-sinusoidal component that occurs
early in the Short High state. The flux of the quasi-sinusoid subsequently
declines. The early flux increase of the quasi-sinusoid may simply be part of
an overall increase in flux as the line-of-sight passes through a decreasing
density of coronal disk material at the start of the High state.} The
most complicated pulses are apparent at the beginning of the High states
with a subsequent disappearance of features as the High state progresses.
Nature would have to conspire so that increases in the amplitude of pulse 
profile components occur only during the Low state. 

Petterson et al. (1991) have qualitatively presented a time dependent
disk obscuration model for the pulse evolution that relies on obscuration 
provided by ``flaps'' of matter at the points 
where a tilted accretion disk meets the neutron star's magnetosphere. These
``flaps'' rotate at the pulse period and move in pulse phase as the tilted disk 
precesses. It is not clear whether the physics of disk-magnetosphere coupling 
will produce obscuring flaps of this type, but in any case the model has some 
qualitative problems when confronted by the observed pulse profiles. 
During the Short High state, the interpulse flux for 
the ``flaps'' model is 
larger than during the Main High state (compare Figs.~2b and 3b of 
Petterson et al. 1991). The {\it Ginga} observations show that the soft peak
flux is {\it the same or smaller} during the Short High state than that of the 
Main High state interpulse with which it is identified (see figure 5). Similar 
behavior in the {\it Exosat} observations can also be seen in Fig.~1 of 
Petterson et al. (1991). During the Short High state the ``flaps'' model 
predicts that the main pulse flux should decrease due to increasing ``flap'' 
obscuration, while the interpulse flux should increase, contrary to the 
{\it Ginga} observations. In the Main High state, the main pulse 
flux and the interpulse flux should also change in an opposing manner according
to the ``flaps'' model, but both fluxes are observed to decrease.   

\section{Pulse evolution resulting from a resolved occultation of the neutron 
star by a tilted, precessing accretion disk}

The apparent stability and repeating nature of the pulse evolution may be
the result of a resolved occultation of the pulse emission
region by the inner edge of a tilted, precessing accretion disk.
Two different occultation sequences, and hence two different sets of pulse
shape changes will occur, as the observer sees the inner 
disk edge sweeping alternately `upwards' and `downwards' across the pulse 
emitting region and these events will be $\sim 180\deg$ apart in disk 
precession phase. The pulse shape changes observed during both
High state types require the ``scale height'' of the disk (or more generally
the ``occulter'')
to be roughly the size of the dominant pulse emitting region i.e. a few neutron
star radii, which will then naturally produce pulse shape changes associated
with decreases in X-ray flux. To pursue the occultation idea further a simple 
geometric model is developed and qualitatively compared to the observations.
The model proposed below is a considerably refined version of the model
originally proposed by Bai (1981).

A disk occultation model requires at least two components. A model for the 
tilted and precessing disk itself is necessary, and a model for the pulsar 
emission geometry. In the simplest approximation, the pulsar beams do not 
depend on the azimuth of the disk and the disk simply occults the pulse 
emitting region. This is probably not entirely true since the disk is coupled 
to the neutron star through the magnetosphere and the azimuthal progression of 
a tilted disk may cause some changes in the accretion column and hence the 
pulse shape. However, these complexities will be ignored for now and the 
beams from the pulsar are assumed to be decoupled from the disk.
With the choice of a physical location for the beam emission region with 
respect to the neutron star, the apparent spatial location of each beam 
component as seen by the observer can be computed. Likewise, the occulting 
disk may have also have a complex geometry due to interaction with the 
magnetosphere, among other effects, but we will assume a simple planar disk 
shape. The sequence of pulse shape changes can be modeled by sweeping the 
disk over the pulse emission region and comparing the predicted pulse shape 
changes with the observations. 

The least obscured pulse should occur early in the Main High state
after the pulsar emerges from behind the outer disk rim.
The main and interpulse profiles are clearly asymmetric about their maxima 
at this time but we will provisionally make several assumptions to simplify
the modeling process. We will first assume that the beam is axisymmetric, the 
magnetic field is purely dipole and that identical beam emission regions exist 
at the ends of an axis that is close to but not necessarily identical with the 
magnetic dipole axis. We attribute the {\it hard central peak} of the main 
pulse to a pencil beam directed along the beam axis and the softer flanking 
shoulders to a surrounding fan beam (similar to Kahabka 1987). The interpulse 
is produced as the edge of the fan beam emanating from the antipodal magnetic 
pole grazes the observer's line-of-sight. 

We assume gaussian intensity profiles for the beams since Kahabka (1987) has 
shown that the pulse profile of Her X-1 can be well fit with a small number of
gaussian components and this is confirmed by fitting the Ginga profiles. A 
simple gaussian beam model for both the fan and pencil beams is used and is 
given by:
\begin{equation}
I(\theta)=I_{pen}\exp(-\theta^2/\sigma_{p}^2)+
I_{fan}\exp((\theta_{cone}-\theta)^2/\sigma_{f}^2)+I_{D}+I_{E}+I_{Low}
\end{equation}
where $\theta$ is the angle from the beam axis, $I_{pen}$ and $I_{fan}$ are the 
pencil and fan beam amplitudes and $\sigma_{p}$ and $\sigma_{f}$ are the beam 
widths. The fan beam opening angle is given by $\theta_{cone}$. Three
constant flux components are present, two due to magnetospheric emission 
($I_{D}$ and $I_{E}$) and one due to Low state coronal emission $I_{Low}$.
Values for
the beam components are given in Table 1. The soft two shoulder component
in the Main High state main pulse results from a cut of the line-of-sight 
across the two edges of the fan beam ``cone'', while the interpulse results
from a grazing cut along the edge of the ``cone''. The large difference in the 
amplitudes of the main pulse and the interpulse require the line-of-sight to 
be $\sim20\deg-40\deg$ from the neutron star's rotational equator. Since 
previous optical and X-ray observations show that the observer is offset by 
$5-10\deg$ relative to the binary plane (Gerend \& Boynton 1976; 
Middleditch 1983; Deeter et al. 1991) the neutron star rotation axis must 
be inclined by $\sim10\deg-50\deg$ to the orbital axis. 

To model the physical location of the primary pulsar beams requires some 
additional assumptions. Theoretically, beam models have been divided into
slab and column geometries depending on the physical mechanism assumed to
decelerate the infalling plasma. Column models assume that either a 
radiation pressure shock or a collisionless shock lies above the neutron star
surface and decelerates the flow. Brainerd \& M\'esz\'aros (1991) have shown 
that the radiation pressure for the luminosity of the Her X-1 is too weak to 
significantly decelerate the infalling plasma. It thus seems prudent 
to assume a slab geometry for the magnetic polar cap of Her X-1. 
In the slab model the infalling plasma is decelerated at the neutron star 
surface and the emitting region is a thin cap only a few meters in height 
(M\'esz\'aros \& Nagel 1985). In more complex models, X-ray radiation
emitted from the foot of the accretion column is backscattered as it rises
through the accretion column (e.g. Brainerd \& M\'esz\'aros 1991). Thus the 
beam consists of direct emission from the polar cap and a backscattered 
component.  

We will first consider a simple pulsar model in which a single emitting point 
lying on the neutron star surface is chosen to approximate the {\it beam} 
emission region. The {\it pulse} emission region will have a width of 
$\le 2 R_{ns}$, where $R_{ns}$ is the neutron star radius. An observer will 
see the emitting point sweep 
out an ellipse as the neutron star rotates with the surface blocking the 
emitting point 
for a portion of the rotation period. We will refer to this beam geometry as a 
``direct fan beam'' geometry. The beam pattern originating from the emitting
point consists of a pencil beam surrounded by a concentric fan beam as in
equation 2 (see figure 8, top panels). This was the same beam pattern and 
emission geometry used in the free precession model in section 5. 

We will also consider a second beam/emission model in which the fan beam 
emission is produced by backscattered radiation from the accretion column that 
is focused around the neutron star and beamed in the antipodal direction, 
a ``reversed fan beam'' geometry.  
This beaming configuration will produce a similar pulse profile but the 
spatial locations of the pulse components will be significantly altered 
(see figure 8, bottom panels). The fan beam components will now be observed 
to originate at some distance above the neutron star surface and 
$\theta_{cone}>90\deg$. The {\it interpulse} will now be emitted by the same 
magnetic pole that produces the {\it hard central peak} while the 
{\it soft shoulders} of the main pulse come from the opposite pole. A similar 
fan and pencil beam configuration has been discussed theoretically by Brainerd 
\& M\'esz\'aros (1991). In their model, a fan beam is produced from magnetic 
polar cap radiation that is preferentially backscattered by the incoming 
accretion flow and then gravitationally focused around the neutron star. The 
accretion column is calculated to be optically thin to Thomson scattering 
while the fan beam photons are produced by cyclotron resonance scattering. 
Only photons at energies less than or equal to the surface cyclotron frequency 
will be scattered in the accretion column. Support for this pulse profile
interpretation is provided by the observed cyclotron absorption feature
in Her X-1, which reaches a maximum absorption depth at the pulse phase of 
the hard central peak and by the disappearance of the main pulse shoulders 
above the $\sim38$ keV cyclotron line energy (Soong et al. (1990b)).

The location of the reversed fan beam emission was modeled by assuming 
the emission occurred from a point at a height of $1.5\ R_{ns}$. This
is the cyclotron scattering height of 10 keV photons for a surface
field strength corresponding to a cyclotron line energy of 40 keV,
a simple dipole field and a neutron mass of $1.4\ \Msun$. Softer photons
will scatter from higher up and harder photons from lower down so the
apparent emission region location is energy dependent. The pulses displayed
in figure 8 approximately model the Main High state pulse profile observed 
in the {\it Ginga} $9.3 - 14$ keV band. Gravitational lightbending of the 
photon trajectories causes the photons to appear to be emitted from a region 
higher above the neutron star surface than for the case when no lightbending
is present. We illustrate this effect in figure 9 where trajectories 
are calculated using the equations taken from Brainerd \& M\'esz\'aros 
(1991) and Riffert \& M\'esz\'aros (1989). A reversed fan beam with an 
opening angle of $\theta_{cone}=140 \deg$ with respect to the accretion 
column and a profile with a half width of $\sigma_{f}=20 \deg$ was used to 
calculate the nonattenuated fan beam intensity since this was a good
approximation to the beam pattern shown in figure 9. The location of the 
fan beam emitting point was calculated using a sinusoidal approximation to 
the photon impact parameter for the case of lightbending in figure 9. 
 
The disk is modeled as an infinite plane with a circular hole centered
on the neutron star. The circular hole has a radius $R_{inner}$ and
a gaussian density profile in the direction perpendicular to the disk plane 
characteristic of an 
$\alpha-$disk of Shakura \& Sunyaev (1973). The vertical disk density 
profile is described by:
\begin{equation}
\rho(z)=\rho_0 \exp(-z^2/\sigma_{d}^2)
\end{equation} 
where $z$ is the vertical distance above the inner disk midplane,
and requires the specification of the free parameters $\rho_0$, the 
midplane density of the disk and $\sigma_{d}$, the disk ``scale height''. 
To model the occultation, the disk is assumed to be simply a linear
edge with a minimum distance $R_{occ}$ from the neutron star as seen by the 
observer.  

The disk selectively obscures the pulse emitting region. 
When the observer's line of sight lies at an angle 
$\theta_{occ}$ above the disk midplane the disk edge will appear to be 
at a distance 
$R_{occ}=R_{inner}\sin(\theta_{occ})$ from the neutron star.
The optical depth of the disk material will be caused by Thomson scattering
due to the complete ionization of hydrogen and helium in the inner disk
region. The optical depth will therefore follow the density distribution
and is assumed to scale as:
\begin{equation}
\tau(z)=\tau_{disk}\exp(-z^2/\sigma_d^2)
\end{equation}
where $\sigma_d$
is the optical depth scale height, and $\tau_{disk}$ is the optical
depth at $z$=0.0. 

The total attenuation caused by the disk for any emitting point depends 
only on its height above (or below) the disk midplane and the angle 
$\theta_{occ}$.
The observer's line-of-sight to the emitting point will pass through a range of 
heights above the disk plane, so the total optical depth must be found by 
integrating along this path. 
The total optical depth is calculated using:
\begin{equation}
\tau(z_0)=0.5(\tau_{mid}/{\sin\theta_{occ}})\int_{z_0}^{\infty}
\exp(-{z \over\sigma_d \cos\theta_{occ}} )^2({dz \over \cos\theta_{occ} }) 
\end{equation}
where $z_0$ is the height above the disk midplane where the ray connecting
the observer and the emitting point intersects the inner disk edge.
The extinction to the emitting point is then given by:
\begin{equation} 
A(z_0)=e^{-\tau(z_0)} 
\end{equation}
The effect of the twisted disk is taken into account by having the optical
depth increase to infinity as $z_0$ becomes increasingly negative (see
figure 2).

The geometric orientation of the neutron star and the inner disk are determined
by tilt and azimuth angles. The tilt is specified with
respect to an axis that will be identified as (but is not required to be) 
the stellar binary axis. Reasonable values for the disk tilt lie in the range 
of $10-20\deg$ in order to fit the overall High state light curve with a 
tilted, twisted, counter-precessing disk (see figure 2). If one assumes
the disk edge is cutting across the face of the neutron star at 35-day
phases 0.23 and 0.58 (from observing the pulse evolution) then a tilt of
$11 \deg$ can be derived.

\subsection{Comparison with Observations}

\subsubsection{Direct Fan Beam}

Can this simple model reproduce, qualitatively, the features observed in 
the Main High state pulse evolution? Figure 8 (top panel)
 shows the approximate spatial 
locations of the pencil beam and fan beam emission regions for the direct
fan beam model. The neutron star rotation is prograde in figure 8 
so a counter-precessing inner disk edge will sweep across the neutron star
face from right-to-left in the figure. 
We define case 1 disk orientation as illustrated in the top panel of figure
8, that is, the disk covers the neutron star from top-to-bottom as well
as right-to-left. 
Case 2 disk orientation is defined as the case that the disk covers
the neutron star face from right-to-left, bottom-to-top, as illustrated 
in the lower panel of figure 8. A straightforward consideration of the
twisted disk geometry shows that during a full  35-day precession period,
case 1 and case 2 both occur, separated by one-half a precession period.

The relatively long period during the 
Main High state when the pulse profile shows only small changes, simply means 
that the line-of-sight to the pulsar is far from the obscuring inner disk edge. 
As the disk edge approaches the pulsar, it is clear from figure 8 (top panel), 
that in 
both case 1 and case 2, the {\it trailing} shoulder (Bb) of the main pulse will 
disappear before the pencil beam (C) or the leading shoulder (Ba).
In case 2, the {\it interpulse} (A) will disappear before 
the pencil beam (C) and in case 1 they disappear at nearly the same time. 
Neither case reproduces the pulse evolution behavior observed in either 
High state.

Reorienting the neutron star spin axis about the line-of-sight 
will alter the occultation sequences. If the upper spin pole in figure 8
is tipped toward the right (toward the oncoming disk edge) 
then case 1 will produce a pulse 
profile sequence in which the order of disappearance will be 1) the 
leading and trailing soft shoulders (Ba and Bb), 2) the hard central peak (C),
 3) the interpulse (A).
Case 1 does not resemble the observed Main High state pulse evolution sequence. 
But it can
can reproduce the Short High state sequence if the emergence from the 
turn-on only reveals the occultation {\it after} the disappearance of 
the two soft shoulders. 
If case 1 gives the Short High state sequence then case 2 must give the Main 
High state sequence.
However, in case 2 the trailing shoulder (Bb) disappears first, then
the hard central peak (C) will 
disappear, both before the leading shoulder (Ba), contrary to the observations.

One can consider all the different possible combinations of neutron star spin
and motions of the disk edge. This has been done and the results (of
whether a satisfactory occultation sequence is obtained) are summarized
in Table 2. Only two of the eight possibilities result in occultation
sequences in the correct order for both the Main and Short High states. One 
requires a reverse neutron star spin in one case and a prograde disk 
precession in the other case. 
Is a retrograde rotation of the neutron star or a
prograde precession of the disk possible? The long term spinup trend observed 
in Her X-1 (Nagase 1989) and the frequency behavior of optical pulsations 
from the lobes of the companion star HZ Her (Middleditch \& Nelson 1976) both
strongly argue for a prograde pulsar spin. Likewise, the preeclipse dip 
recurrence period and the predominance of
integer combinations of the {\it sum} of the orbital and 35-day frequency
in the power density spectrum of the optical light curve 
strongly indicate a {\it counter}-precession for the accretion disk 
(Deeter et al. 1976; Crosa and Boynton 1980).
Thus we discard the direct fan beam model, since it cannot give the correct
occultation sequence and keep prograde spin and retrograde precession.

We also note that those occultation sequences from the direct fan beam model,
even though in the correct order, have a difficulty. For the Short High state 
the observed sequence starts with a weak hard pulse (C) but has no evidence
whatsoever for any trace of the leading (Ba) or trailing (Bb) shoulders.
This is very hard to achieve in the direct fan beam model since the hard
pulse (C) and soft shoulders (Ba, Bb) are produced in directly adjacent 
locations. 
The above qualitative difficulties in matching the observed sequence of pulse 
profile changes can be resolved using a ``reversed'' fan beam, as we show 
next.

\subsubsection{Reversed Fan Beam}
  
Both the observed Main and Short High state pulse evolution 
can be reproduced qualitatively 
if the neutron star is tilted as shown in the lower panels of figure 8.
Case 1 (disk tilted as in the upper panel) 
reproduces the Main High state decay of the {\it leading} soft 
shoulder (Ba) followed by the {\it hard central peak} (C) 
and the {\it trailing} soft shoulder (Bb). The Short High state 
evolution is produced by the case 2 disk orientation as shown in the lower
panel of figure 8. The observed evolution requires complete occultation of the 
soft shoulders (Ba, Bb) of the main pulse before the Short High state turn-on.
This can be readily accomplished since the shoulders (Ba, Bb) are from
regions well separated from the location of the {\it hard central peak} (C). 
The shorter length of the Short High state and the required initial partial 
occultation of the pulse emitting region can be produced by the same offset 
of the observer's line-of-sight from the binary plane. In figure 10a and 10b,
a disk occultation and the resulting pulse profile changes are illustrated
using the model disk and beam profiles described earlier. The parameters
used are displayed in table 1.

The Main High state pulse evolution is modeled in figure 10a. 
The leading edge decay of both the fan beam and the interpulse occurs 
as the disk edge sweeps across the neutron star face. The actual timing of 
the decay of the soft shoulders and the interpulse put important constraints
on the apparent tilt of the neutron star relative to the disk edge, which
is also quite sensitive to the actual locations of the emitting
regions. For the model shown in figure 10a, the interpulse decays away a
bit early relative to the soft shoulders compared to the observations. In
addition the decay of the trailing soft shoulder is delayed compared to
the observations since the model disk edge appears to be too sharp. 
However, making the disk edge fuzzier will tend to wash out the Short
High state sequence, assuming a fixed inner disk ring tilt and radius.
This model predicts that the pencil beam ({\it hard central peak}, C), 
should show a trailing edge decay, but observing this effect requires a 
known pulse ephemeris during this decay phase. Lastly, as the disk edge cuts 
across the neutron star face, short term variations in the pulse profile
should occur in addition to the longer term systematic changes, due to 
variations in the disk opacity as material is accreted onto the neutron star. 

The Short High state pulse evolution is modeled in figure 10b.
The disk motion will cause the pencil beam ({\it small hard peak}, C) to decay 
away first followed by the interpulse (A). The more edge-on view of the inner 
disk plane at the start of the Short High state will lengthen the traversal of
the disk edge across the neutron star face relative to the Main High state
and this is consistent with the slow disappearance 
of the interpulse ({\it soft peak}, A) relative to the same event during the 
Main High state. In contrast to the Main High state, no significant leading 
edge decay of the pulse components is predicted during the Short High state and 
none is observed. The Short High state must begin with the accretion disk
partially obscuring the pulse emission region to account for both the
different pulse shape and the lower Short High state flux. The 
placement of the inner disk edge at the turn-on is probably somewhat variable
due to changes in the total disk twist and elevation so this model predicts
that the Short High state pulse profile should be quite variable as well at 
the start of the turn-on. Depending on the exact placement of the inner disk 
edge, the {\it small hard peak} (C) may be larger, smaller, or completely 
attenuated compared to the interpulse in figure 6. 
Likewise, the disk edge may be advanced enough that the interpulse is gone as 
well at the turn-on.

The evolution of the quasi-sinusoidal components is modeled by the occultation
of emission regions D and E in figure 10.
The deepening of the pre-interpulse minima early in the Main High state 
and the persistence of the quasi-sinusoidal pulse after the disappearance 
of the main pulse suggests an occultation of an emission region much larger 
in size than the neutron star. Radiation scattered from the inflowing 
magnetospheric material many neutron star radii above the surface may produce 
this emission. If we assume 
1) the magnetospheric flow is optically thin to Thomson scattering as in 
the Brainerd \& M\'esz\'aros model and 2) that accretion onto a magnetic pole 
occurs over a range of magnetic azimuth less than $180\deg$, then two equal and 
constant flux pulse components may be produced by scattering in the accretion 
flows. The locus of scattering points illuminated by the hard pencil 
beam (C) over a neutron star rotation period are indicated by D and E
in the rightmost panels of figure 8. D and E are much further away from the
neutron star than the emission components A, Ba, Bb and C (middle column 
of panels in figure 8): the small circle between D and E indicates the
neutron star surface.
Since the emission from E is due to the magnetospheric flow onto the 
the opposite pole 
on the neutron star from the flow producing the emission at D,
the instantaneous location of scattered emission on E will 
be $180\deg$ different in pulse phase from the instantaneous location of 
the emission on D. D and E are still close enough to the neutron
star that light travel time delays are negligible.
In the absence of any occultation, the instantaneous
scattering from D and E are visible at all pulse phases so the total
emission from D and E appears unmodulated.
A third, small, constant-flux contribution should also be present from 
the much larger accretion disk corona that produces the Low state flux.

The Main and Short High state quasi-sinusoidal pulse evolution and the
preinterpulse deepening can together be interpreted as an occultation of the 
X-ray illuminated, magnetospheric flow (regions D and E). 
Both the upper and lower
magnetospheric flows will sweep out cones as the neutron star rotates, which
cross the ``plane of the sky'' at pulse phases 0.25 and 0.75. 
For the neutron star orientation show in figure 8 (lower panels),
a deepening is predicted in the pulse profile during the Main High state 
near pulse phase 0.25 as the 
outer portion of the upper magnetospheric flow 
rotates behind the oncoming disk edge. 
The {\it widening} will continue as the 
occultation progresses but the {\it deepening} will stop since the flux at 
pulse phase 0.25 will then be produced mostly by the unocculted, 
antipodal, magnetospheric flow. 

Near the middle of the 
Main High state, a second deepening is predicted at pulse phase 0.75 as the 
disk begins to occult the lower  
magnetospheric flow. At the Short High state turn-on, the lower magnetospheric 
flow will already be occulted, accounting for
the $\sim 50\%$ drop in flux observed in the ``quasi-sinusoid''. 
The constant flux produced by the upper magnetospheric flow will now be
preferentially occulted near pulse phase 0.75 as the occultation
progresses, producing a dip there, and an $\sim 180\deg$ phase shift 
between the two High state ``quasi-sinusoids''. 
The dip should widen as the Short High state progresses. 
The initial increase and 
then decrease of the overall quasi-sinusoidal flux occurs simply because the
line-of-sight is moving away from the outer disk edge and through decreasing
accretion disk coronal density as the inner disk occultation progresses.

A comparison of the model in figure 10 with the observations readily shows
that while many features of the quasi-sinusoidal evolution can be
modeled, there is a problem with component E. At the end of the Main High
state, the quasi-sinusoid peaks at pulse phase 0.75, whereas component
D is near minima, implying that it is out-of-phase by 0.5. The Short High
state provides no constraint on E since it is completely occulted. The
``fix'' to the simple model needed to produce the asymmetry in the soft 
shoulder peaks should also affect the location of pulse component E and may 
solve this problem. In any case, using two rotating rings to model the 
quasi-sinusoids is probably a gross simplification of the actual situation 
since the interaction geometry of the accretion column with the disk should be 
quite complex. In addition, another weakly pulsed component may be present due
to scattering or reemission from the inner disk edge. This may be the cause
of the low energy sinusoid that appears to be in phase
with the higher energy quasi-sinusoid (see e.g. McCray et al. 1982). We note 
that the pulse evolution of
the soft energy quasi-sinusoid ($<\ 1$ kev) is at present unknown and that its 
observation may provide valuable clues about the inner disk. 
Recent BeppoSAX observations of a Short High state flux decline
(Oosterbroek et al. 2000) discovered an increase in relative absorption that 
can be explained by assuming seperate scattering and absorption regions. This 
is in qualitative agreement with a gradual inner disk occultation of an extended
scattering region associated with the accretion column. 

The radius of the inner disk edge can be estimated from the duration
of the Main High state occultation event. Let $R_d$ be the inner disk radius 
and $R_e$ the radius of the pulse emitting region. From the reversed beam 
model $R_e$ is about $\sim 2 \; R_{ns}$ for the region emitting the main pulse.
The velocity of the disk edge is given approximately by 
$V_d=R_d \omega_d \sin \theta_t$ where $\omega_d$ is the angular velocity
of the disk precession (equal to ${2 \pi \over P_{35}}$) and $\theta_t$
is the tilt angle of the inner disk. The inner disk radius can now be
estimated from:
$$ R_d={2R_e \over T_{occ} \omega_d \sin \theta_t} $$
where $T_{occ}$ is the duration of the occultation of the main 
pulse emitting region. From the April 1989 Main High state, 
$T_{occ}\approx 3$ days. The inner disk tilt angle is estimated to be 
between $10-20\deg$. These parameter values produce an estimate of 
$R_d\approx 20-40 \; R_{ns} $. This is much smaller than the corotation
radius of $157 R_{ns}$ where the orbital angular velocity equals that
of the neutron star,   
assuming a 12 km neutron star radius and $M_{NS} = 1.3 \Msun$.     

In conclusion, the reverse fan beam model with prograde neutron star spin 
and retrograde disk precession naturally accounts for the major features
of the observed Main and Short High state pulse evolution. It also
accounts for a number of other features, as described above, e.g.
the $180\deg$ phase shift between the two High state ``quasi-sinusoids''. 

\section{Discussion}

The model proposed here for the pulse evolution cycle in Her X-1
consists of an occultation of the neutron star by the inner edge of a tilted
and precessing disk. The leading edge decay of the main and interpulses during
the Main High state preclude a beam geometry consisting of a pencil beam
surrounded by a fan beam and emitting from the neutron star surface. However, 
reversing the fan beam so that is emitted
in the antipodal direction with respect to the pencil beam and at some 
distance above the stellar surface 
allows the leading edge decay to be
reproduced by an occultation in a natural fashion. The observed Short High
state evolution pattern then arises naturally with this geometry. 

Cyclotron resonant scattering in the accretion column 
is an attractive mechanism for producing a pencil beam surrounded
by an reversed fan beam.
The energy of the cyclotron resonance is directly proportional to the
local magnetic field strength and this decreases with altitude above the 
neutron star surface. When the energy of an upward traveling photon equals 
the local cyclotron energy, scattering will occur. This creates a natural 
energy 
dependent filtering process for photons emitted in the accretion cap and
divides the beam into three components. Hard photons (especially those
above the surface cyclotron frequency) will escape in a pencil beam. Softer
photons will be backscattered and gravitationally focused around the neutron
star in an antipodally directed fan beam. Finally, photons scattered from the
accretion column isotropically will produce a constant pulse component. 
The softest photons will scatter from the highest altitudes in the accretion 
column. The neutron star rotation will cause 
the highest altitude emission (D and E in figure 8) to rotate with pulse phase.
The occultation of this high altitude emission 
by the precessing disk edge creates gaps in the constant
emission profile and therefore a ``quasi-sinusoidal'' profile. 
A hard pencil beam (C), a softer fan beam (Ba and Bb)
and a quasi-sinusoidal component (D and E)
qualitatively match the observed energy dependence in the Main High state 
profile. This type of beam model may be applicable to other X-ray pulsars
as well, for example Vela X-1, which displays a pulse profile consisting of
two doubled peaked components at soft X-ray energies, possibly superposed on
a quasi-sinusoidal component, that fills in at harder X-rays in a way quite 
similar to the Her X-1 main pulse (White, Swank \& Holt 1983). In the case of 
Vela X-1, the observer's line-of-sight would have to be located much closer 
to the neutron star spin equator than in Her X-1. 

While an occultation model apparently has many attractive features, a 
roughly $45\deg$ tilt is required between the neutron star spin axis and the 
binary axis of the system. Is this plausible? A large tilt to the neutron star
may have originated in the supernova explosion that gave it birth. The accretion
of matter from the companion will cause angular momentum to be
accreted by the neutron star. Since the direction of the time averaged 
accreted angular momentum is along the binary axis of the system, the neutron 
star's spin axis 
should become coaligned with the binary axis. This event will take some time
and the accreted angular momentum will spin up the neutron star. In fact,
the alignment timescale and the spin-up timescale should be comparable.
The measured spin-up time of Her X-1 is about $10^{5}$ years. This timescale
is the same as that predicted for the entire X-ray emitting phase of
Her X-1 (Savonije 1978). Therefore, if Her X-1 is currently tilted that tilt
will in all likelyhood be maintained throughout the rest of the X-ray
emitting phase. However the important question is what is the {\it current} 
ratio of accreted angular momentum to that at the start of the X-ray 
emitting phase?

Her X-1 may be near the start of the X-ray emitting phase.
The X-ray emitting phase ends when the accretion flow 
becomes great enough to smother the neutron star and prevent pulsations.
Historical optical observations of HZ Her show that the X-ray heating and
hence the mass accretion rate has ceased occasionally for years to 
decade-long periods over the last hundred years (Jones, Forman \& Liller 1973;
Hudec \& Wenzel 1986). 
The current state of mass transfer depends on
the X-ray heating of HZ Her and is inhibited by the X-radiation pressure.
Thus HZ Her must be close to, but not quite, filling its Roche lobe.
This state of affairs is more 
consistent with Her X-1 being at the beginning rather than the end of its 
X-ray phase. If so, then a highly tilted neutron star can plausibly exist in 
the Her X-1 system. Another point in favor of such an interpretation is the 
1.24 second pulse period of Her X-1. This period is typical of the radio pulsar 
population. A period longer than 3 seconds or much shorter than 1 second 
would unequivocally show that Her X-1 has been spun down or spun up by a 
significant history of interaction with circumstellar matter. Finally,
a tilted neutron star should cause a persistent asymmetry in the optical
orbital lightcurve about orbital phase 0.5. No such asymmetry was reported
by Deeter et al. (1976) but later studies of the orbital optical lightcurve
with more data indicate the presence of just such an asymmetry (e.g. 
Voloshina, Lyuti \& Sheffer 1990; Thomas et al. 1983).

The properties required of the inner disk to fit the observed pulse evolution
are a scale height comparable to the neutron star diameter and a small inner
radius ($20-40 R_{ns}$). 
For comparison, the corotation radius 
is $157R_{ns}$ for Her X-1.
The predicted magnetospheric radius, $R_m$, for disk accretion depends on 
assumptions on the boundary conditions. 
Using the known parameters for Her X-1 (Leahy \& Scott 1998), one obtains 
$R_m=3.5\times10^8\alpha^{-2/61}$ cm for the model of Kiraly \& M\'esz\'aros (1988);
$R_m=3.8\times10^8\gamma^{2/7}$ cm for the model of Lamb (1988); and
$R_m=4.9\times10^8$ K cm for the formula of Finger et al. (1996) 
(which has the special case of $K=0.47$ for the model of Ghosh \&
Lamb, 1978).
The model of Aly (1980) for a highly conducting disk, gives 
only a slightly smaller value for the disk inner radius: 
$R_m=3.0\times10^8\alpha^{2/7}\cos(\chi)^{4/7}$ cm (for $\chi$ near $\pi/2$,
the dependence on $\chi$ is $\sin(\chi)^{40/69}$, so the minimum $R_m$ is
$\simeq1/3$ of this for $\chi=0$).
The disk viscosity parameter is $\alpha$, $\gamma$ is defined in 
Lamb (1988), K is a dimensionless parameter, which Finger et al. (1996) find
to be $\simeq 1$ for A0535+26, and $\chi$ is the tilt of the dipole axis 
from the equatorial plane.
In all cases the predicted disk radius greatly exceeds the inner disk
radius required by the occultation model.

Another observation relevant to the magnetospheric radius is the spinup rate of 
Her X-~1. 
It is the smallest among the accretion-powered
pulsars and indicates the slowest rate of net angular momentum accretion.
During the giant outbursts of Be-transients, an X-ray flux vs. spin-up rate 
correlation has been observed in which spin-up occurs at a rate consistent
with the fiducial torque (e.g. Finger et al. (1996); Bildsten et al. (1997); 
Nelson et al. (1997); Scott et al. (1997)). 
The formula for the fiducial accretion torque $N_f$ is given by 
$N_f=\dot M(GM_{X}R_s)^{1/2}$ where $R_s$ is the magnetospheric radius
for spin-up, $\dot M$ is the mass accretion rate and 
$M_{X}$ is the mass of the neutron star. This formula
assumes all the angular momentum of the accreting matter at $R_s$ 
is given to the neutron star.
The known spin-up rate gives a value for $R_s=8.5$ km: clearly too small to
be physical. 
The small disk radius from the occultation model is much larger than this.
This implies that the actual torque on the neutron star must
be smaller than the fiducial torque, to allow a magnetospheric radius
as large as $20-40R_{ns}$. 
In the Ghosh \& Lamb (1978, 1979) model of an aligned rotator, 
a residual portion of the magnetic field is not screened by
currents near the magnetospheric radius and will interact with material orbiting
beyond the corotation radius. The interaction adds a negative torque component 
to the total torque.
A smaller net torque will be exerted if the magnetospheric radius approaches the
corotation radius\footnote{Note that Ghosh and Lamb prefer to call the 
corotation radius the `centrifugal radius' and the magnetospheric radius the 
`corotation radius'.} of the neutron star. 
Thus nearly any spin-up rate can be obtained within a small range of
magnetospheric radii near to but inside the corotation radius. So it is
not surprising that the predicted magnetospheric radius from the Ghosh and
Lamb model for Her X-1 is nearly equal to the corotation radius given above.
However, it is well known that the Ghosh and Lamb model 
is not consistent with the spin behaviour of many X-ray pulsars (e.g.
Bildsten et al. 1997; Nelson et al. 1997). In summary, different models 
give a magnetospheric radius from the spin-up rate of Her X-1 in a
wide range between $R_{ns}$ and $R_c=157R_{ns}$.

How does one account
for an inner disk edge which is at $20-40R_{ns}$?
It is much smaller than the predicted neutron star magnetosphere radius, 
yet may be much larger than the radius predicted from simple spin-up. 
We speculate on what may be the real physical origin of the 
$20-40R_{ns}$ disk inner edge.
One possibility is that the dipole magnetic
field is much smaller than that indicated by the observed X-ray cyclotron
line so that the predicted magnetospheric radius is much smaller. 
The magnetospheric radius decreases from $\sim 200R_{ns}$ to $30R_{ns}$ if the
dipole component of B is a factor of 25 less than deduced from the cyclotron
line. The cyclotron line would be explained as arising in non-dipolar magnetic
fields in the accretion region at the neutron star surface. An emission 
region concentrated in a non-dipolar surface pocket of the field and producing
a pencil beam would likely be difficult to distinguish from a similar emission 
region in the case of a pure dipole. Another explanation has been put forth by 
Baushev \& Bisnovatyi-Kogan (1999) in which a magnetic field of 
$4-6 \times 10^{10}$ G is estimated for Her X-1 from the observed cyclotron 
line energy by assuming a large anisotropy exists in the electron momentum 
distributions parallel and perpendicular to the magnetic field lines. Whether 
either of these possibilities is viable will probably require much more 
research.    

An alternative to a reduced dipole field is a stable, thin inner disk which 
penetrates far into the magnetosphere. However, this seems very unlikely: the 
magnetosphere is very stiff due to the steep gradient of magnetic pressure 
and energy density (as $r^{-6}$) for a dipole field. Models which have the 
disk penetrate as deeply as possible (e.g. Aly 1980), do not have an inner 
radius very much less than other models (tilting the magnetic axis gives a 
reduction at most by a factor 3 in the inner disk radius). So a disk 
penetrating far into the magnetosphere appears to be unfeasible.

Another possibility is that current models for determining the magnetospheric
radius are inadequate. For example, Miller \& Stone (1997) use 
magnetohydrodynamic calculations to show that the Balbus-Hawley
instability  and magnetic braking have dramatic effects on the magnetospheric
boundary. They also result in outflowing winds along field lines opened up
by reconnection. It is quite possible that the small inner disk in Her X-1
is due to such effects. We note that mass outflows are quite likely as there 
is extended X-ray emission from a large corona in Her X-1. 

The model described in this paper is phenomenological and highly idealized.
The purpose was to show that an inner disk occultation can explain the 
observations and has reasonable physical grounds for support. Considerable
refinement of the model remains to be done. The decay of the soft shoulders
of the main pulse occurs earlier in the Main High state than an axisymmetric
beam model predicts. The soft shoulders of the Main High state main pulse
are unequal in amplitude. These problems might indicate an offset 
in the dipole axis.

An occultation of the neutron star by the inner accretion disk explains the 
pulse evolution cycle of Her X-1 in a natural fashion. Many of the details
of the observed evolution can be accounted for by invoking a reversed 
fan beam geometry around a neutron star significantly inclined to the
binary axis of the system. The leading edge decay 
of the main pulse and interpulse during the Main High state is properly
predicted. During the Short High state, the rapid disappearance of the 
small hard peak as well as the slower decay of the soft peak are predicted.
Note that the soft peak shows little decay of either the leading or trailing
edge and this is also predicted by the occultation model. In summary,
most of the curious pulse shape changes observed during the Main and Short 
High states are tied together with a single simple occultation model.
We note that no previous model for the pulse shape changes, including
free precession, has been able to account for the observed details 
of the pulse evolution.

\begin{acknowledgements}
DMS acknowledges John E. Deeter and Paul E. Boynton for useful discussions
and Fumiaki Nagase and Takashi Aoki for assistance while visiting ISAS. 
\end{acknowledgements}

% References

% Figure Captions
\newpage

Fig. 1---Light curves (1--37 keV) for {\it Ginga} LAC observations of
Her X-1 in the April, May and June 1989 Main, Short and Main High states,
respectively. The 35-Day phase is calculated using $P_{35}=34.8534$ days 
($20.5P_{orb}$) and an epoch $T_{0}=48478.6 - 31 P_{35} = 47398.14$ MJD with 
phase 0.0 corresponding to the Main High state turn-on. Eclipse ingress and 
egress are denoted by dashed vertical lines. Ticks and numbers at the top of 
each panel indicate the 35-day phase. Solid vertical lines mark predicted 
turn-on time. The bottommost plot also shows the 20-50 keV pulsed flux 
lightcurve obtained from BATSE monitoring of Her X-1, with the flux scale 
shown on the righthand side. 

Fig. 2---{\it (upper panel)} View of the disk as seen from the neutron star. 
The outermost ring (filled diamonds) is tilted by $20\deg$, the innermost 
ring (hollow diamonds) by $11\deg$ and leads in precession phase by 
$138.6\deg$. 
The elevation of the observer of Her X-1 ($-5\deg$) is marked by the 
horizontal dashed line.
% For LMC X-4, the observer
% would be placed at an elevation of $22\deg$ above the binary plane.
Bold solid vertical lines mark turn-on's.
The observer sees Her X-1 emerge as a point source from the outer disk
rim but an extended source at the inner disk rim.
%After turn-on until roughly the point marked by the vertical dotted line,
%coronal obscuration will be decreasing. The 35-day phase of partial blocking
%of the X-ray emission region during the turn-off is marked by dashed sinusoids.
The vertical dotted line shows the approximate point where the disk cuts across
the neutron star face during the Main High state. 
{\it (Middle panel)} The 1989 1--37 keV Her X-1 light curve observed with 
{\it Ginga}.                                                  
{\it (Bottom panel)} Nineteen ``0.2 turn-on'' 35-day cycles of the 2--12 keV 
RXTE ASM light curve folded at a period of 20.5 $P_{orb}$. 

Fig. 3---{\it (upper left panel)} {\it Ginga} observation of 1-37 keV count rate
during the May 1989 Short High state turn-on. Vertical dashed line marks 
predicted eclipse egress. Solid curve models flux of point source rising through
outer disk edge (see text). Errors are plotted, unless smaller than plotted
point size.  
{\it (lower left panel)} Same for August 1991 Main
High state turn-on. {\it (upper right panel)} Softness ratios of two sets of 
energies during the May 1989 Short High state turn-on. Note that harder flux 
increases first consistent with cold matter absorption.  
{\it (lower right panel)} Same for August 1991 Main High state turn-on.

Fig. 4---The softness ratio for the average Main High state light curve. The 
hard color is the BATSE 20-50 kev pulsed flux and the soft color is the 
RXTE/ASM 2-12 keV flux. The {\it top} panel shows the average softness ratio 
for 0.2 turn-on Main High states, the {\it bottom} panel for 0.7 turn-on 
Main High states. Turn-on is 35-day phase 0. Vertical dashed lines mark 
eclipse ingress/egress boundaries.    

Fig. 5---Sample {\it Ginga} pulse profiles in five energy bands,
after subtracting background and correcting for collimator
transmission. The bands increase in energy from top to
bottom: 1.0--4.6, 4.6--9.3, 9.3--14, 14--23, 23--37 keV. The bottommost
panels display a hardness ratio (9.3--23 keV band divided by 1.0--4.6 keV
band). Pulse features discussed in the text are labeled. 
{\it (a)}---Leftmost panel set. Main High state observation on MJD 47643. 
Total exposure time is 8713 seconds. 
{\it Main pulse} occupies phase interval 0.75--1.25 and the {\it interpulse} 
occupies phase interval 0.3--0.7.                                              
{\it (b)}---Center panel set. Same profiles as in (a), but with close-up
of interpulse. 
{\it (c)}---Rightmost panel set. Short High state observation on MJD 47662,
less than a day after Short High state turn-on. Total exposure time is 10118 
seconds. At same scale as center panel set.

Fig. 6---Time evolution of the pulse profile. Upper right and upper
left panels show evolution during the April 1989 Main High state in
energy bands 1.0-4.6 keV and 9.3-14 keV. The flux of the smallest amplitude 
pulse in each panel is correct while offsets of 50 counts/sec have been 
added between pulses for clarity. The 35d phase of the pulses increase
with decreasing amplitude as: 0.05, 0.162, 0.216, 0.243, 0.245, 0.247,
0.249, 0.250. 
{\it Lower left} and {\it lower right} panels.
Same for the May 1989 Short High state but with offsets of 20 counts/sec
added. The 35d phase increases with decreasing amplitude as 0.590, 0.595,
0.613, 0.641, 0.698.

Fig. 7---Sequence of pulse profiles resulting from a freely precessing neutron 
star with an axisymmetric pencil and fan beam using the parameters determined 
by Kahabka (1987,1989). The pencil beam amplitude is three times the fan beam 
amplitude and both beam components have half-widths of about $20\deg$. One 
full 35-day precession cycle is shown with successive pulses occuring one day 
apart and spaced by one flux unit. On the right is the pulsed flux light curve 
over the precession cycle normalized to a maximum value of 10.

Fig. 8---Top left panel. Model pulse comprised of identical beam configurations
at each pole consisting of a central pencil beam (C) and surrounding by a 
fan beam (B) with an opening angle of $40^\circ$. Three constant components 
are present, two produced by isotropic emission high in the accretion column 
(D,E) and one from coronal emission. Top center. Emission locations of  
pulse components A, B and C. Location of disk edge for 35-day phase 0.23 
shown moving from right-to-left, top-to-bottom across figure (case 1).  
Top right. Larger view showing emission locations of pulse components
D and E. Bottom left. Same as above but with reversed fan beam (opening 
angle of $140^\circ$) and light bending taken into account. Bottom center. 
The `A' and `B' components are now emitted when the accretion column is on 
the ``back'' side of the neutron star with respect to the observer. 
Location of disk edge for 35-day phase 0.58 shown moving from right-to-left, 
bottom-to-top across figure (case 2).  

Fig. 9---Top panel. Photon trajectories for $10$ keV photons backscattered
from a cyclotron resonance at a height of $1.5 R_{ns}$ above a $1.4\ \Msun$ 
neutron star. The scattering height assumes a surface magnetic field strength 
corresponding to a $40$ keV cyclotron line and a simple dipole field.    
Bottom panel. The photon impact parameter and apparent emission angle for
the case of no light bending (diamonds) and with light bending (triangles).
Distance units in Schwarzschild radii.

Fig. 10a---Model of Main High state pulse evolution. Left panel sets shows
a sequence of pulse profiles corresponding to the figure 8b, reversed fan
beam case. Middle panel set shows closeup of pulse emission region and
gradual covering by disk. Rightmost panel set shows larger view of same 
region. 35-day phase is shown at far right.  

Fig. 10b---Model of Short High state pulse evolution. Left panel sets shows
a sequence of pulse profiles corresponding to the figure 8b, reversed fan
beam case. Middle panel set shows closeup of pulse emission region and
gradual covering by disk. Rightmost panel set shows larger view of same 
region. 35-day phase is shown at far right.  

%%% Tables

% Table 1
\newpage

\makeatletter
\def\jnl@aj{AJ}
\ifx\revtex@jnl\jnl@aj\let\tablebreak=\nl\fi
\makeatother

% From here on, the file contains tabular data as an author might
% prepare it.

\begin{deluxetable}{lrrrcrrrr}
\tablewidth{33pc}
\tablecaption{Disk occultation model parameters}
\tablehead{
\colhead{Parameter} & \colhead{Value} &\colhead{Symbol} }
\startdata          

Observer inclination & $85 \deg$ & \nl
LOS-rotation axis angle & $61 \deg$ & \nl
Magnetic axis - spin axis angle & $48 \deg$ & \nl
spin axis - binary axis angle & $52 \deg$ & \nl
NS azimuth & $45.5 \deg$ & \nl
Pencil beam scale factor & 2.75 & $I_{pen}$ \nl
Half-width of pencil beam & $18 \deg$ & $\sigma_{p}$ \nl
Fan beam scale factor & 1 & $I_{fan}$ \nl
Opening angle of direct fan beam & $40 \deg$ & $\theta_{cone}$ \nl
Opening angle of reversed fan beam & $140 \deg$ & $\theta_{cone}$ \nl
Half-width of fan beam & $20 \deg$ & $\sigma_{f}$ \nl
Effective height of fan beam emission point & $2 R_{ns}$ & \nl
Inner disk radius & $30.5 R_{ns}$ & \nl
Accretion disk tilt & $11 \deg$ & \nl
Optical depth at disk atmosphere base & 30 & $\tau_{disk}$ \nl
Disk atmosphere e-folding length & $1 R_{ns}$ & $\sigma_d$ \nl
Quasi-sinusoidal emission point height & $15 R_{ns}$ \nl  
Quasi-sinusoidal scale factor (D=E) & 0.1 & $I_{D,E}$ \nl
Low state emission scale factor & 0.1 & $I_{Low}$ \nl

\enddata
\end{deluxetable}                                 

% Table 2
\newpage

\makeatletter
\def\jnl@aj{AJ}
\ifx\revtex@jnl\jnl@aj\let\tablebreak=\nl\fi
\makeatother

% From here on, the file contains tabular data as an author might
% prepare it.

\begin{deluxetable}{lrrrcrrrr}
\tablewidth{33pc}
\tablecaption{direct fan beam disk occultation scenarios}
\tablehead{
\colhead{Spin} & \colhead{Tilt} & \colhead{Precession} & \colhead{Okay?} }
\startdata          

prograde   & left   & retrograde & no  \nl
prograde   & right  & retrograde & no  \nl
prograde   & left   & prograde   & yes \nl
prograde   & right  & prograde   & no  \nl
retrograde & right  & prograde   & no  \nl
retrograde & left   & prograde   & no  \nl
retrograde & left   & retrograde & no  \nl
retrograde & right  & retrograde & yes \nl

\enddata
\end{deluxetable}                                 
                             
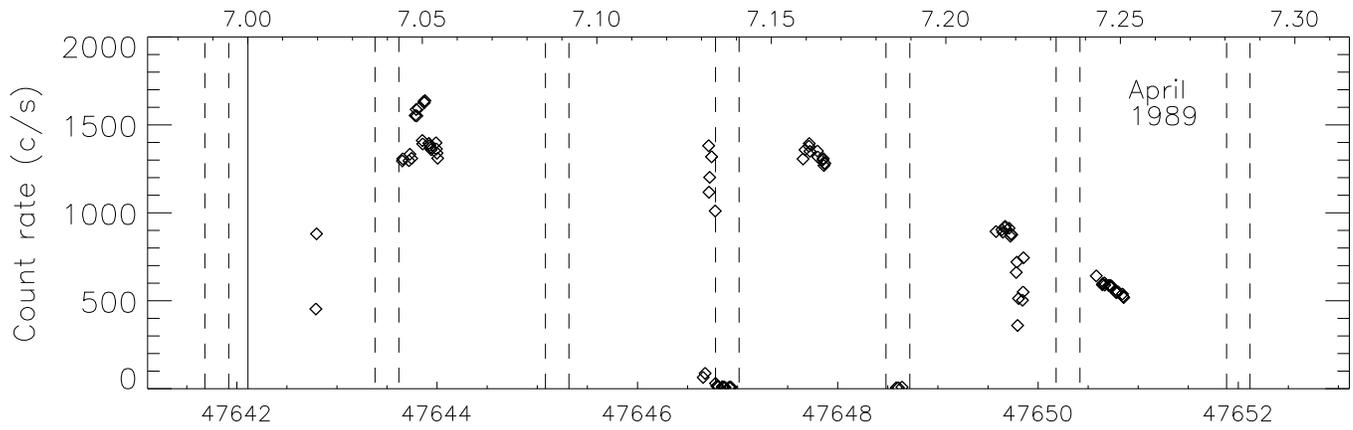
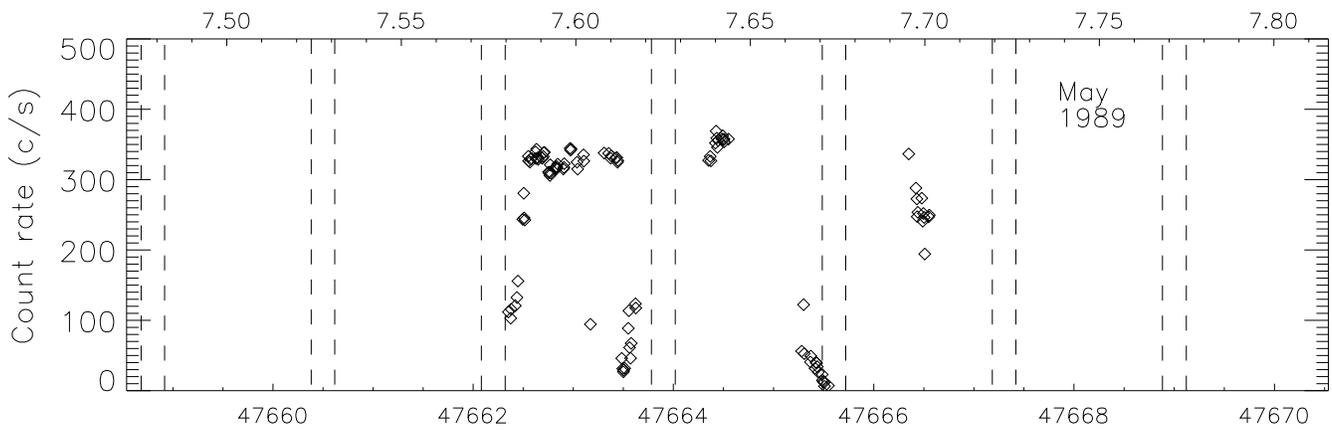
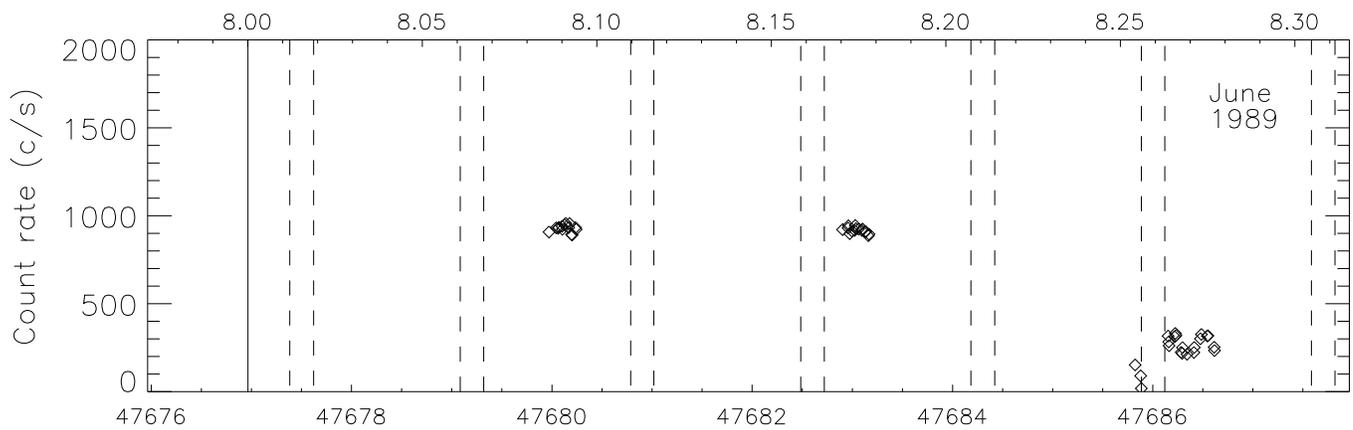
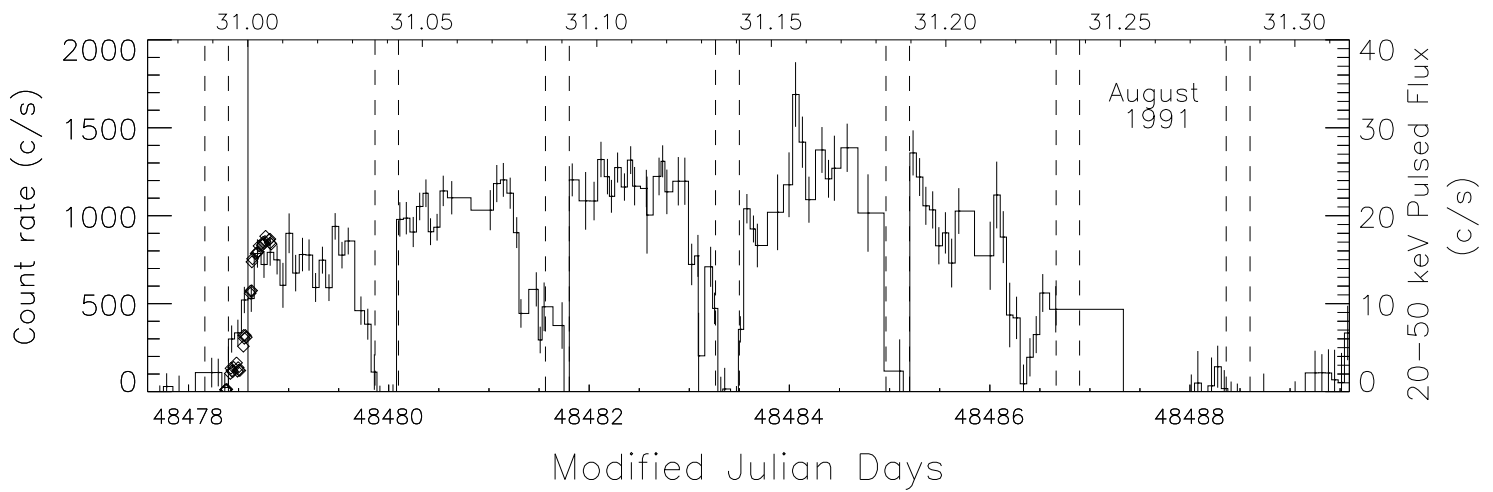
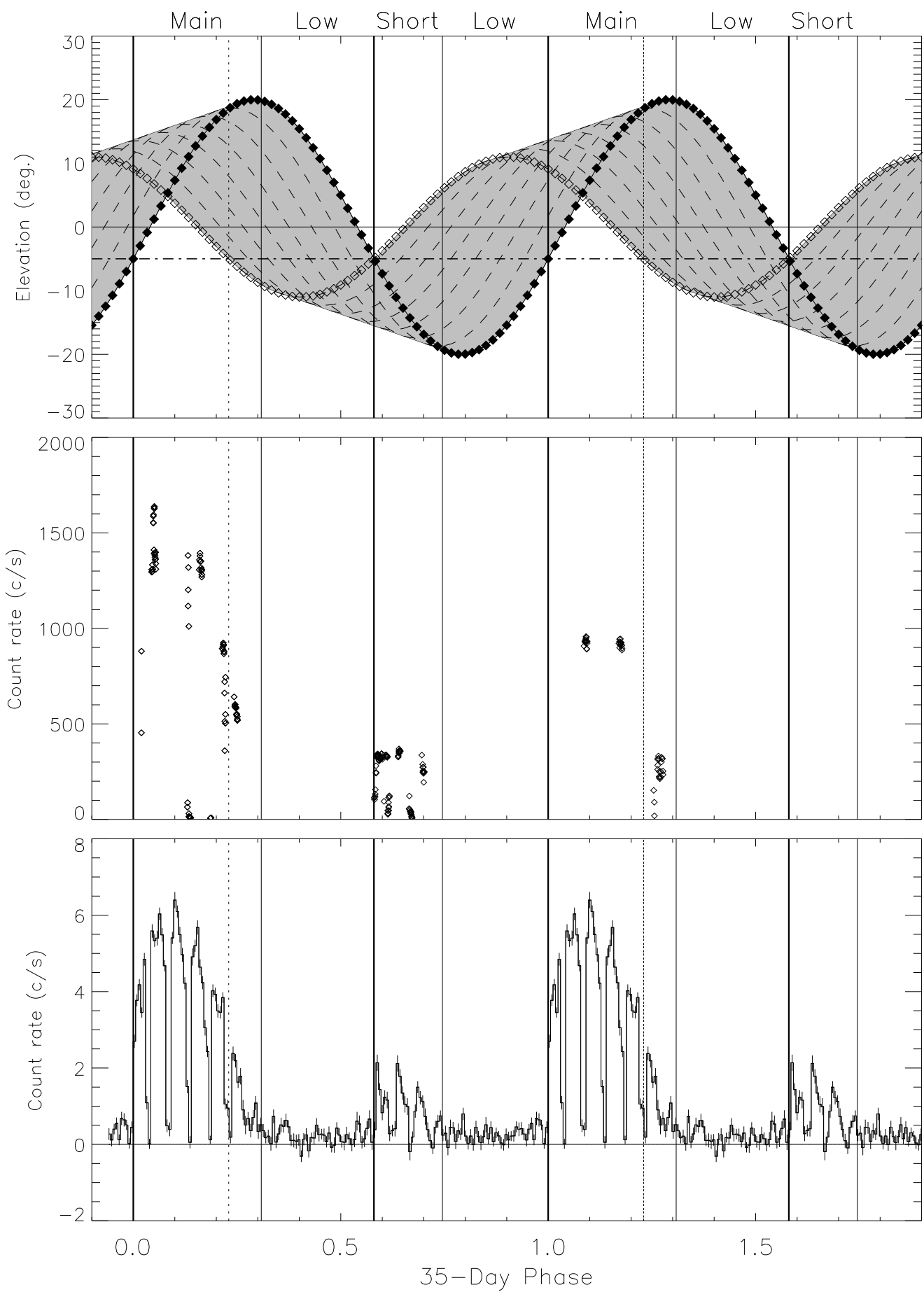
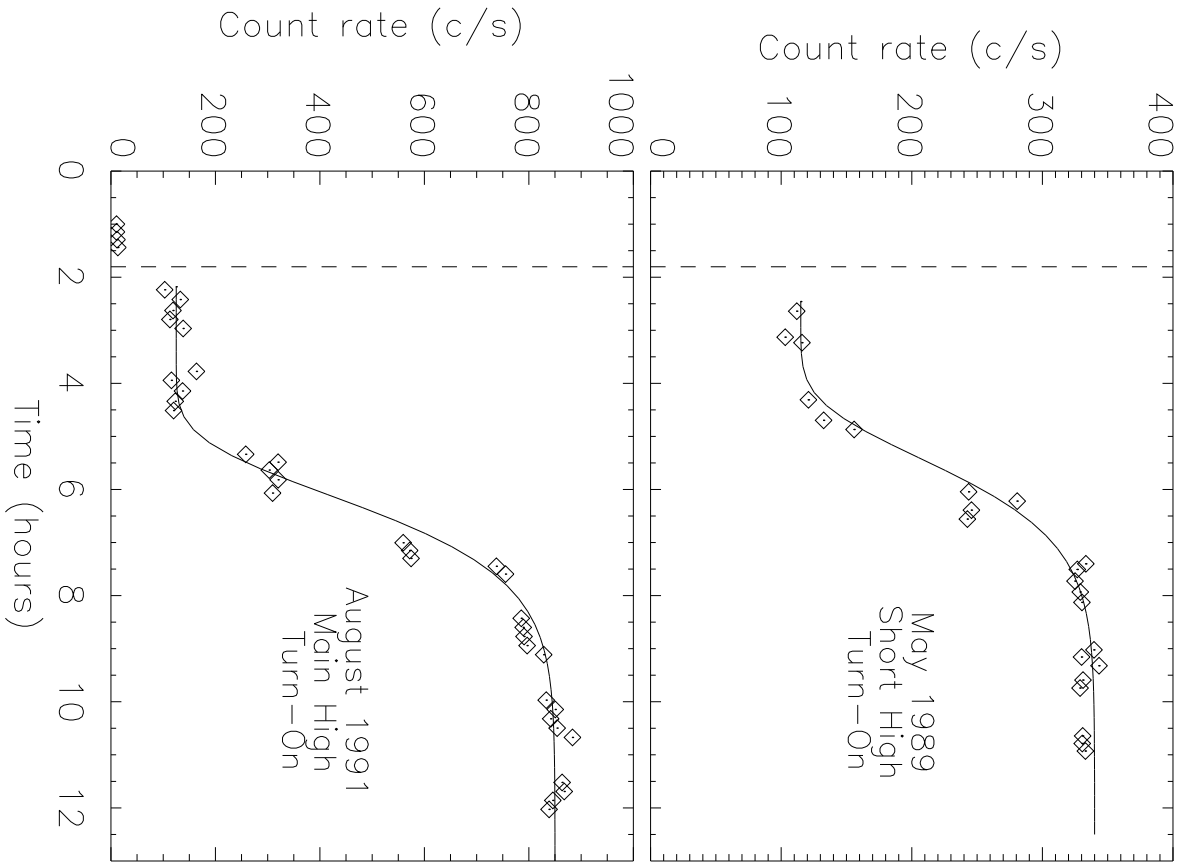
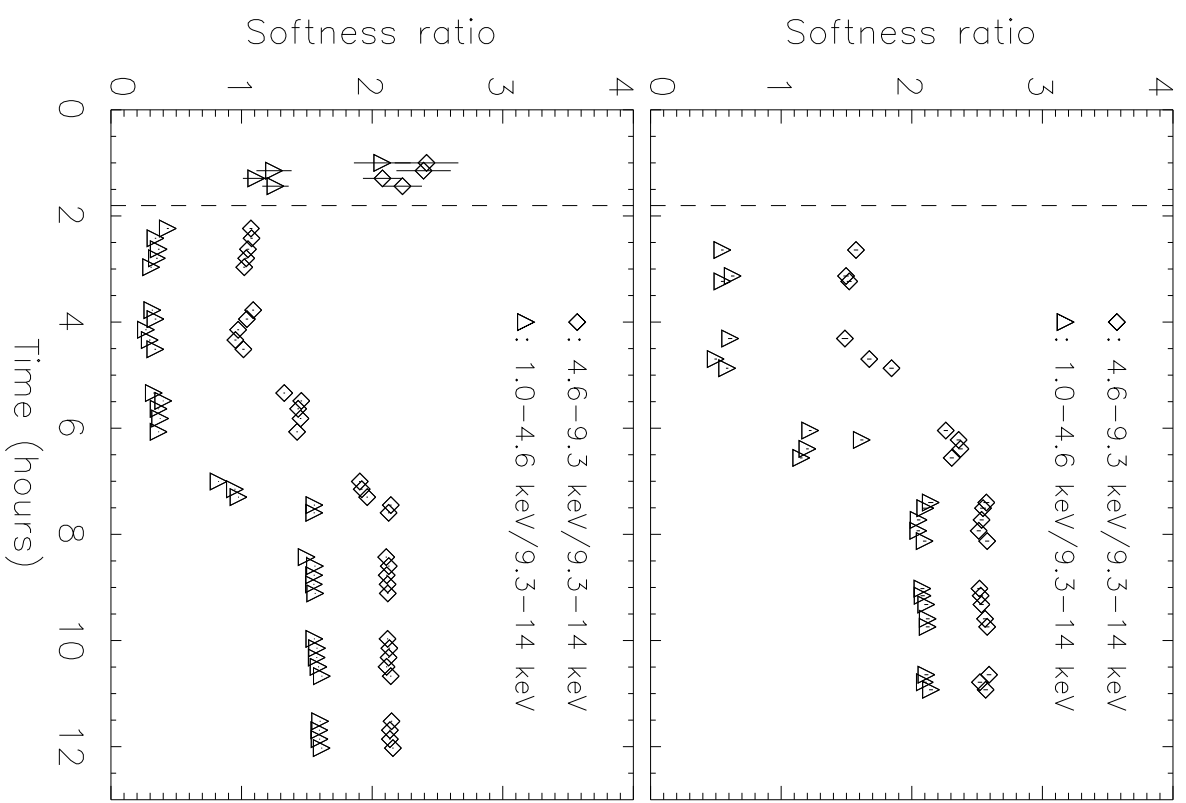
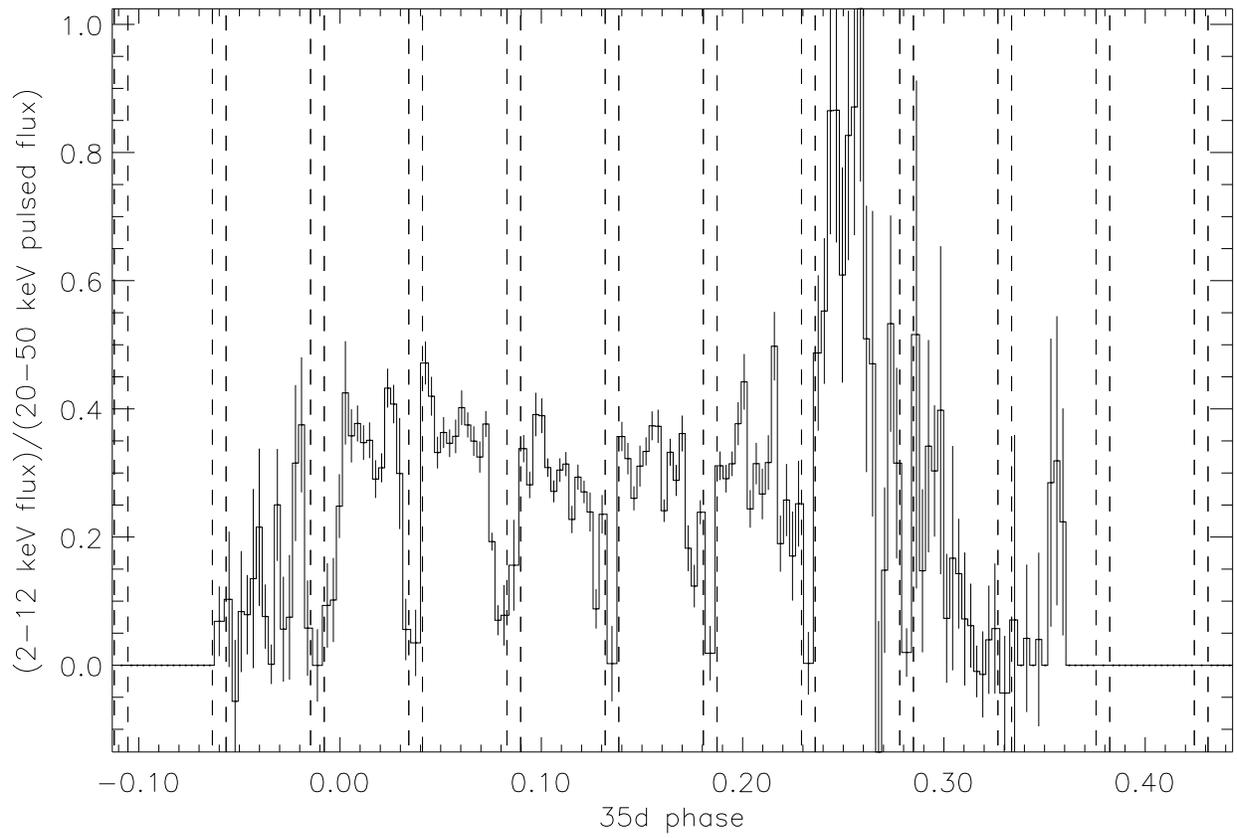
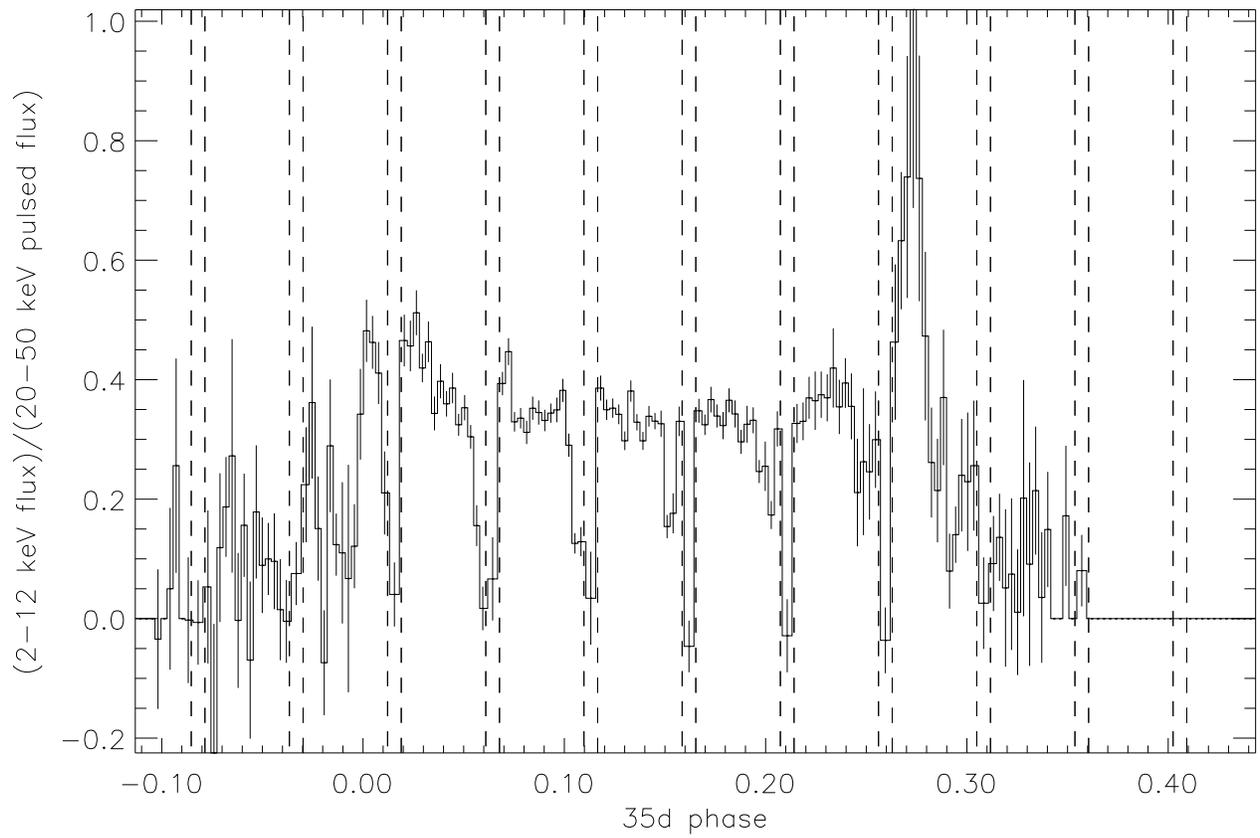
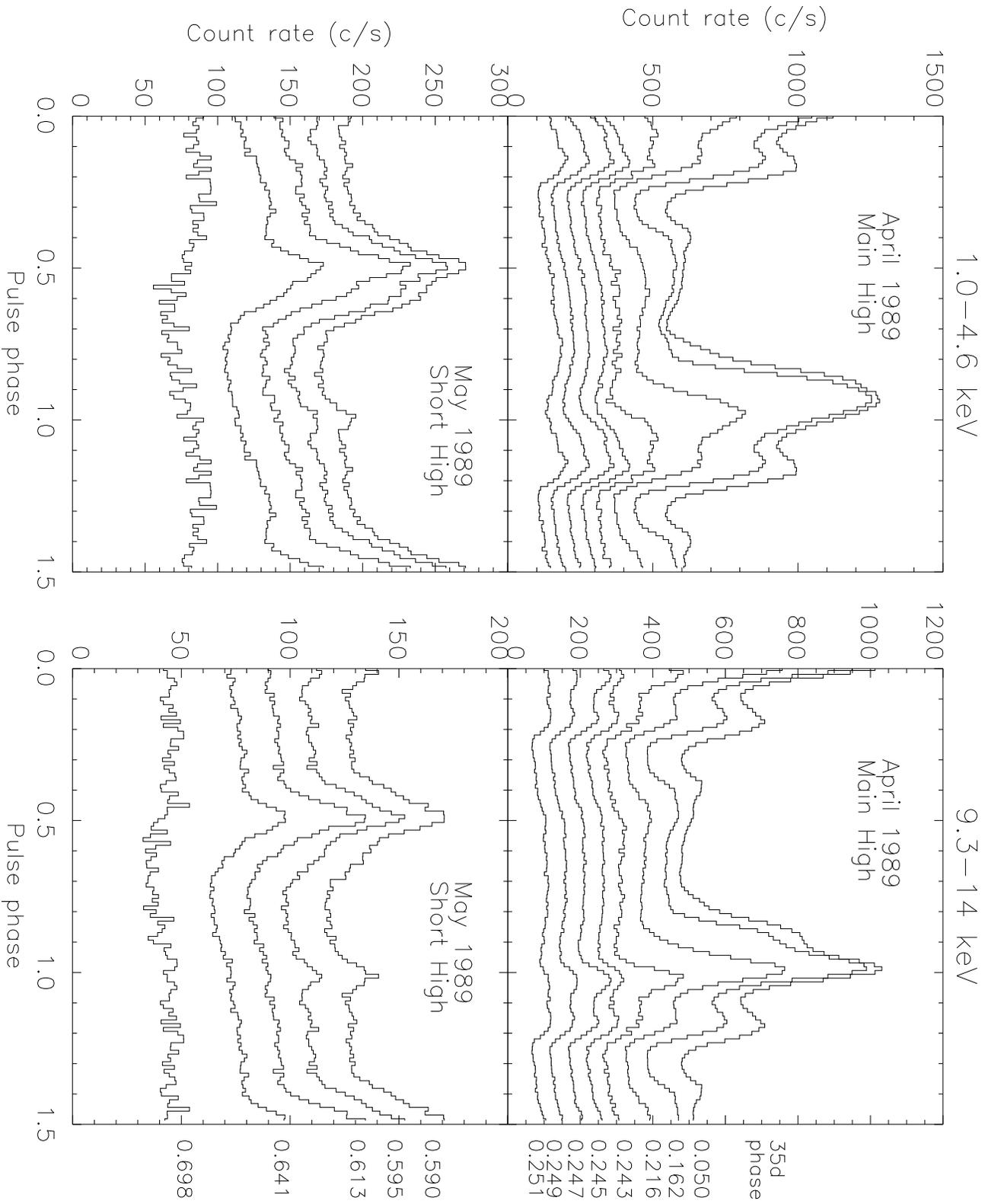
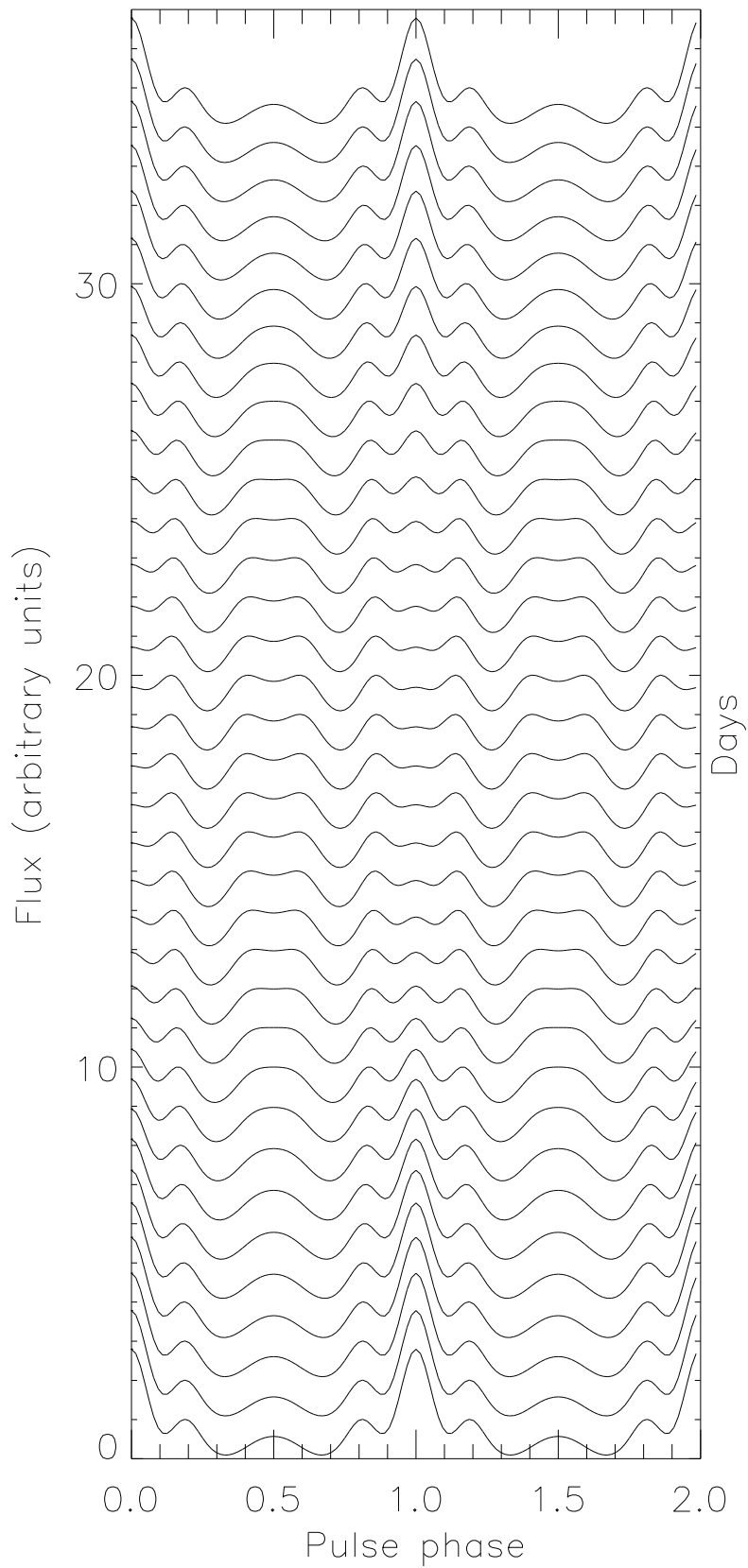
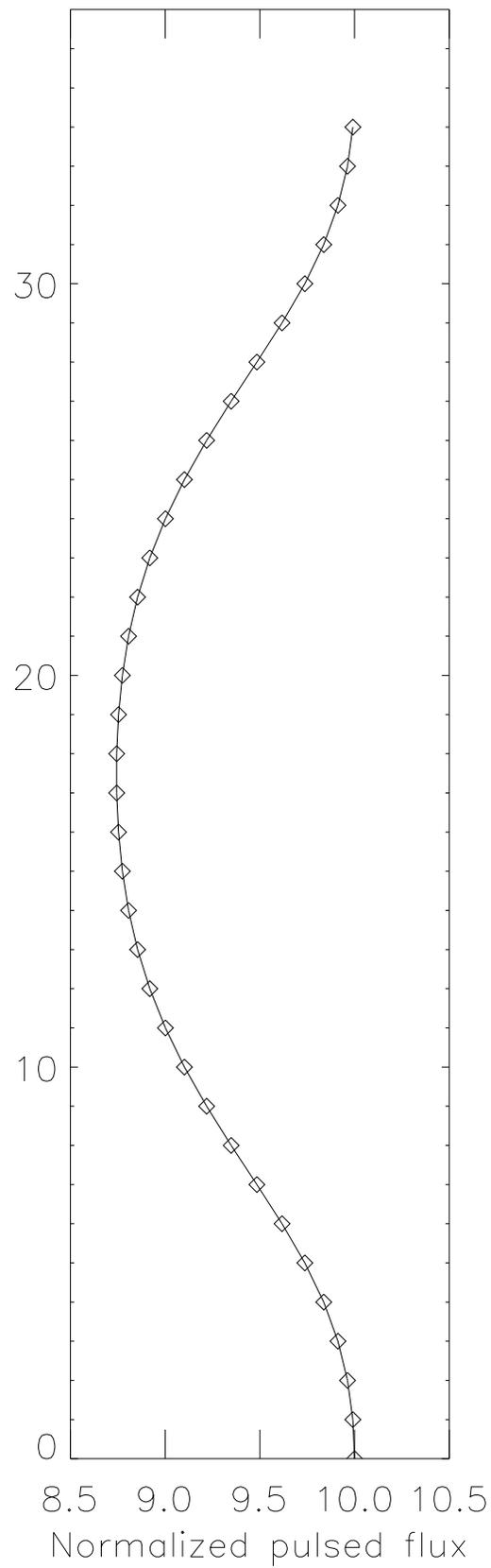
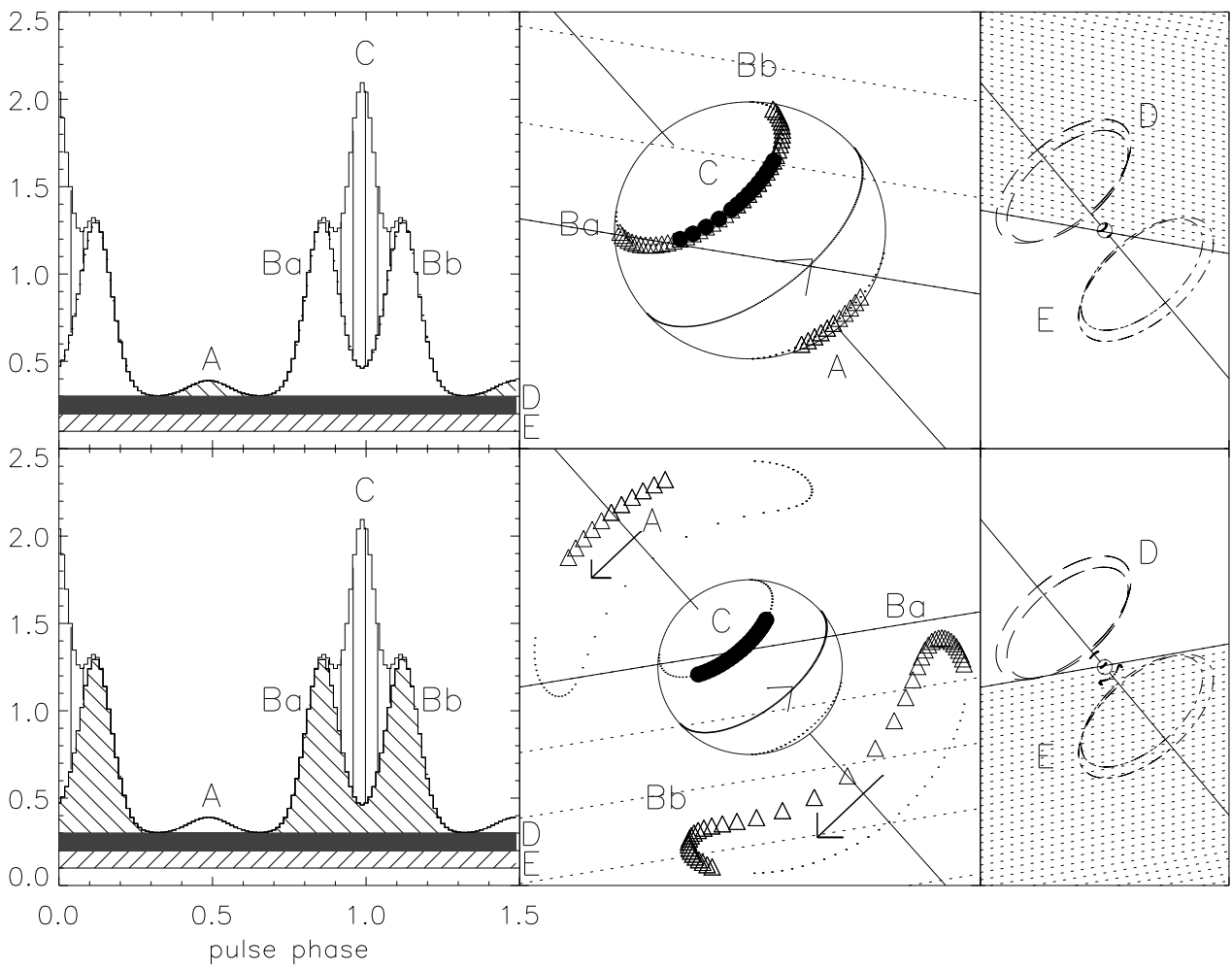
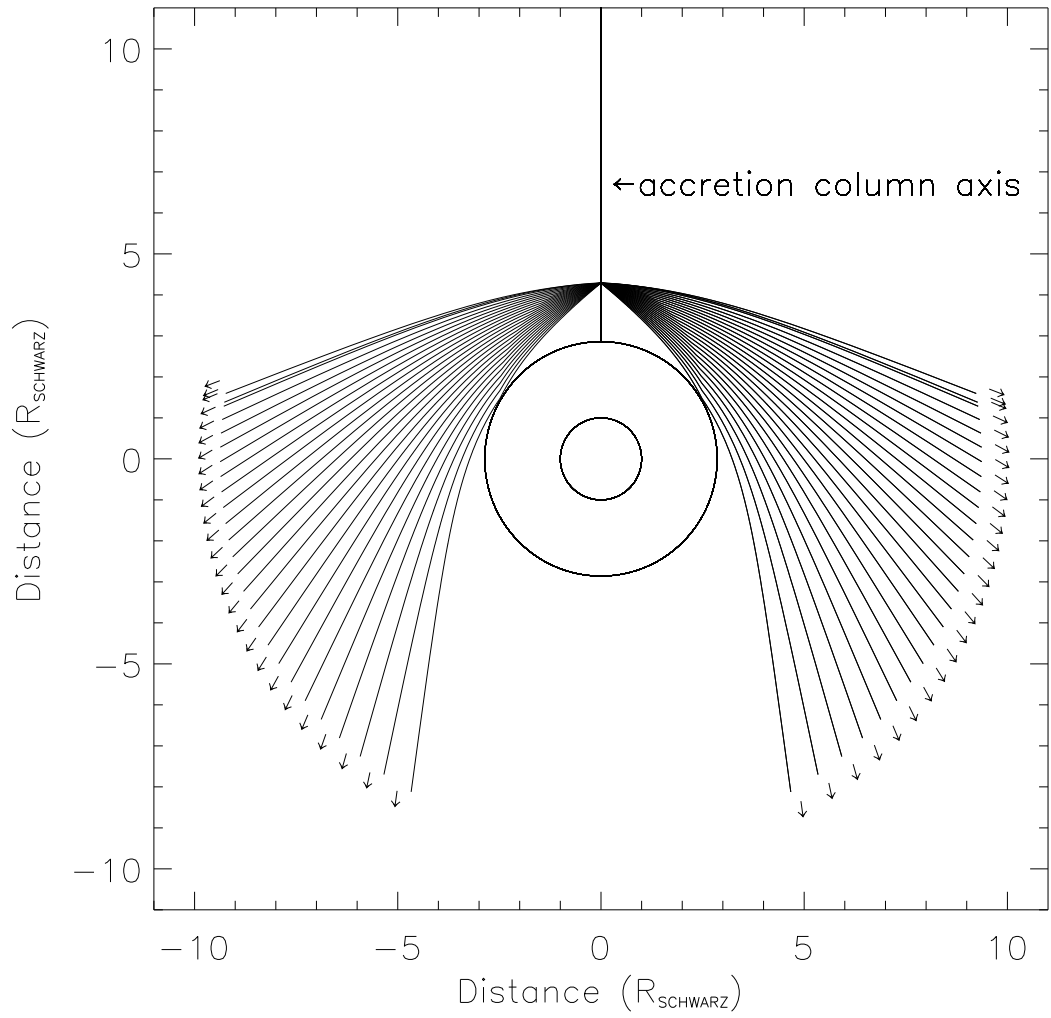
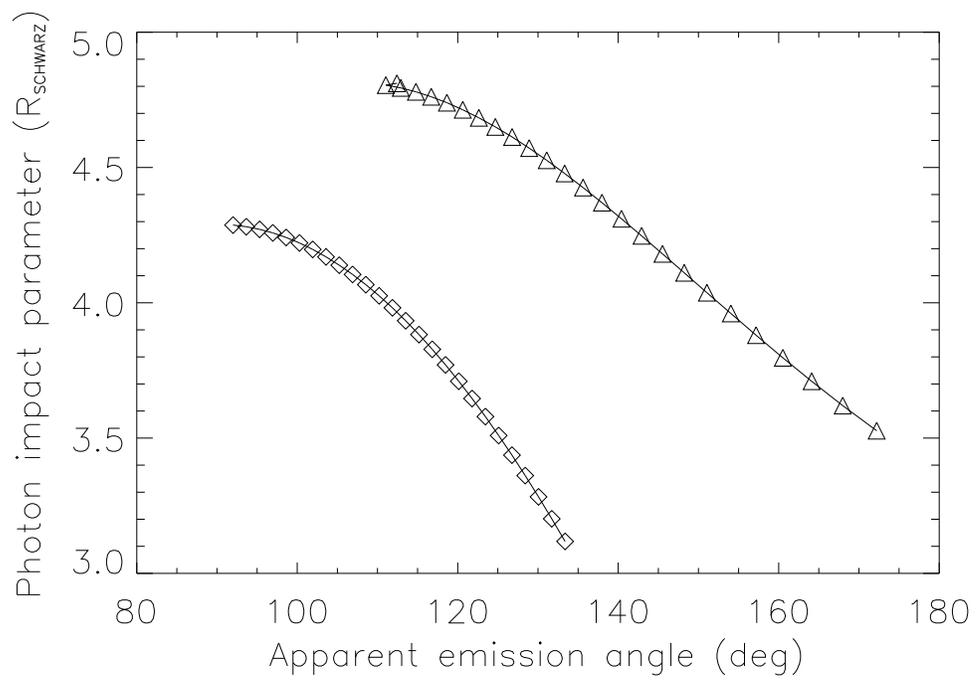
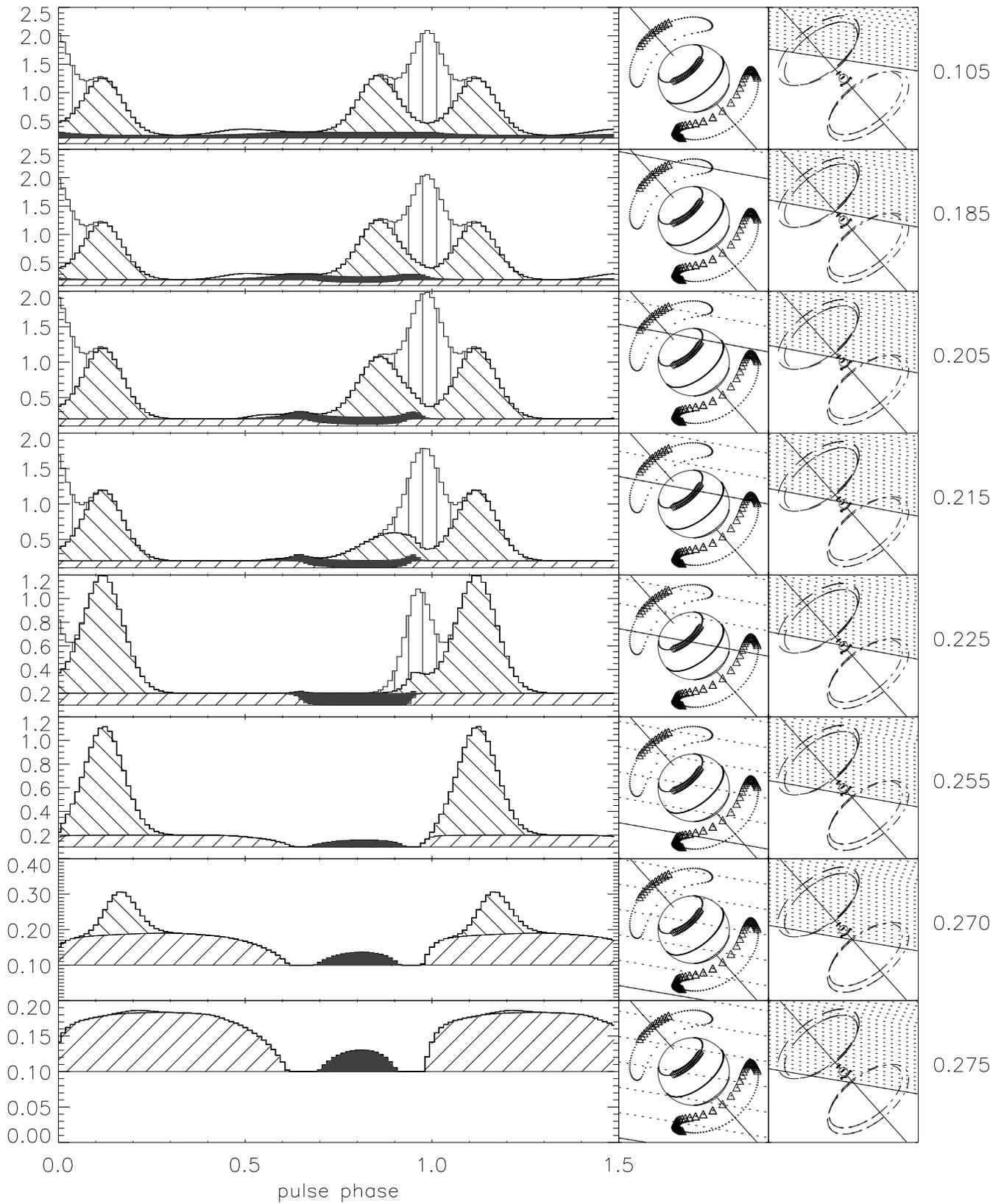
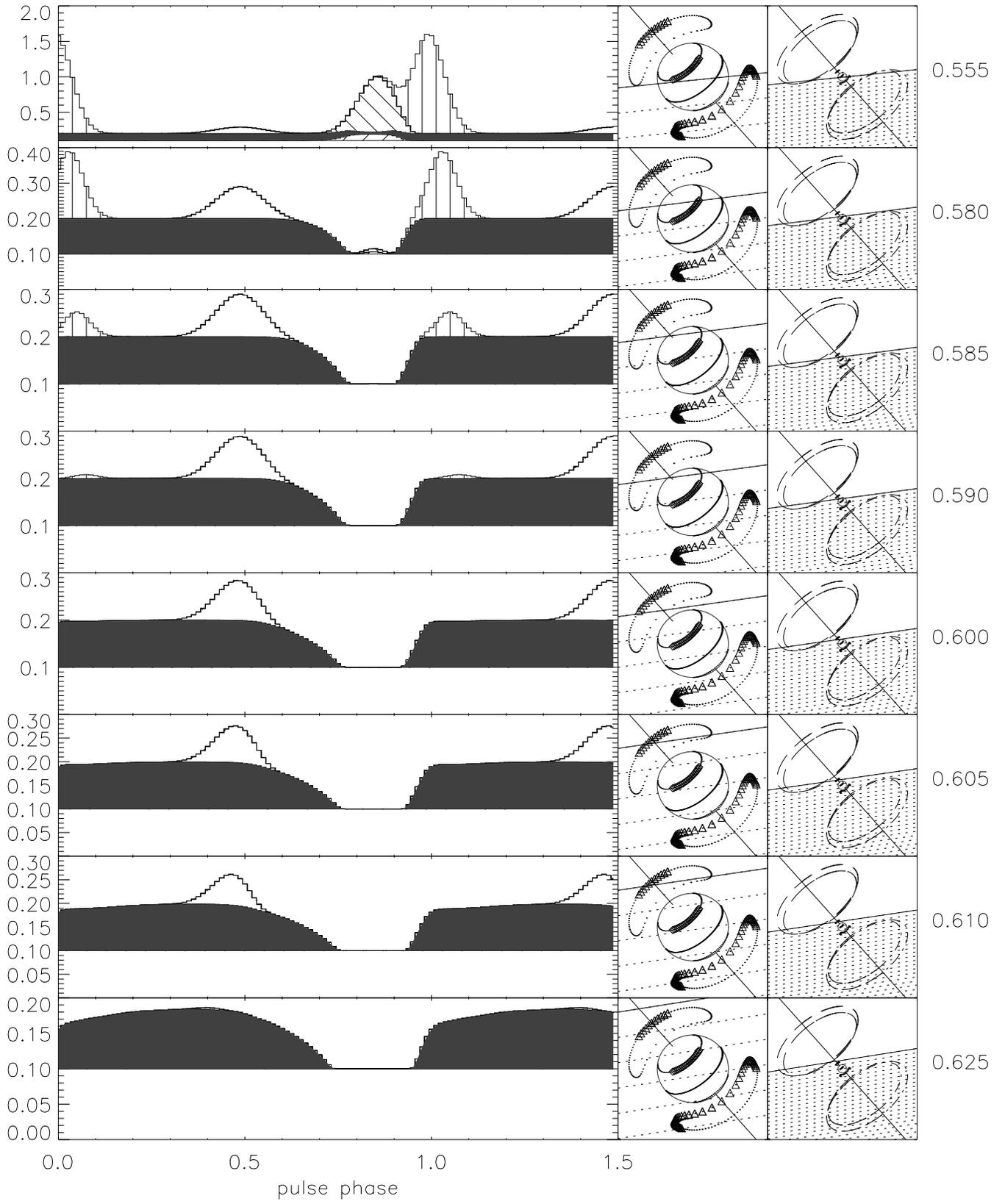
\end{document}